\documentclass[aps,pra,reprint,superscriptaddress,a4paper]{revtex4-1}
\usepackage{amsmath,amssymb}
\usepackage[utf8]{inputenc}
\usepackage{graphicx}
\usepackage{xcolor}
\usepackage{hyperref}
\hypersetup{colorlinks=true,citecolor={blue},linkcolor={blue},urlcolor={blue}}
\usepackage{braket}
\usepackage{bm}
\usepackage{enumitem}
\usepackage[pass]{geometry}

\usepackage[separate-uncertainty=true,multi-part-units=single,group-separator={,}]{siunitx}
\DeclareSIUnit[per-mode=symbol,per-symbol=p]{\ueV}{\micro\electronvolt}
\DeclareSIUnit[per-mode=symbol,per-symbol=p]{\um}{\micro\meter}

\newcommand{\half}{\textstyle{\frac{1}{2}}}

\graphicspath{{./figs/}}
\usepackage{float}

\newcommand{\capind}[1]{(#1)}
\newcommand{\captit}[1]{#1}

\begin{document}

\title{Spin order and phase transitions in chains of polariton condensates}

\author{H. Ohadi}
\email{ho278@cam.ac.uk}
\affiliation{Department of Physics, Cavendish Laboratory, University of Cambridge, Cambridge CB3 0HE, United Kingdom}
\author{A. J. Ramsay}
\affiliation{Hitachi Cambridge Laboratory, Hitachi Europe Ltd., Cambridge CB3 0HE, UK}
\author{H. Sigurdsson}
\affiliation{Science Institute, University of Iceland, Dunhagi-3, IS-107 Reykjavik, Iceland}
\author{Y. del Valle-Inclan Redondo}
\affiliation{Department of Physics, Cavendish Laboratory, University of Cambridge, Cambridge CB3 0HE, United Kingdom}
\author{S. I. Tsintzos}
\author{Z. Hatzopoulos}
\affiliation{FORTH, Institute of Electronic Structure and Laser, 71110 Heraklion, Crete, Greece}
\author{T. C. H. Liew}
\affiliation{School of Physical and Mathematical Sciences, Nanyang Technological University 637371, Singapore}
\author{I. A. Shelykh}
\affiliation{Science Institute, University of Iceland, Dunhagi-3, IS-107 Reykjavik, Iceland}
\affiliation{ITMO University, St.\ Petersburg 197101, Russia}
\author{Y. G. Rubo}
\affiliation{Instituto de Energ\'{\i}as Renovables, Universidad Nacional Aut\'onoma de M\'exico, Temixco, Morelos, 62580, Mexico}
\affiliation{Center for Theoretical Physics of Complex Systems, Institute for Basic Science (IBS), Daejeon 34051, Republic of Korea}
\author{P. G. Savvidis}
\affiliation{FORTH, Institute of Electronic Structure and Laser, 71110 Heraklion, Crete, Greece}
\affiliation{ITMO University, St.\ Petersburg 197101, Russia}
\affiliation{Department of Materials Science and Technology, University of Crete, 71003 Heraklion, Crete, Greece}
\author{J. J. Baumberg}
\email{jjb12@cam.ac.uk}
\affiliation{Department of Physics, Cavendish Laboratory, University of Cambridge, Cambridge CB3 0HE, United Kingdom}

\begin{abstract}
    We demonstrate that multiply-coupled spinor polariton condensates can be
    optically tuned through a sequence of spin-ordered phases by changing the
    coupling strength between nearest neighbors. For closed 4-condensate
    chains these phases span from ferromagnetic (FM) to antiferromagnetic
    (AFM), separated by an unexpected crossover phase. This crossover phase is composed
    of alternating FM-AFM bonds. For larger 8 condensate chains, we show the critical role of spatial inhomogeneities and
		demonstrate a scheme to overcome them and prepare any desired spin state.
    Our observations thus demonstrate a fully controllable non-equilibrium spin lattice.
\end{abstract}

\maketitle

Spin models, such as the Ising model, have been very successful in describing a
wide range of condensed matter phenomena~\cite{baxter_exactly_2016}. In
addition, these models can be mapped to real-world optimization
problems~\cite{kirkpatrick_optimization_1983,lucas_ising_2014}, for example in
transport scheduling, artificial intelligence, and financial portfolio
optimization~\cite{hayashi_accelerator_2016,Yamaoka_AIIsing_2016}. Consequently,
there is a growing interest in building controlled spin lattices both to study
computationally complex spin systems such as spin
glasses~\cite{barahona_computational_1982}, but also as a potential computing
architecture~\cite{buluta_quantum_2009,georgescu_quantum_2014}. Several systems
have been explored, including ultracold atoms \cite{bloch_quantum_2012},
degenerate optical parametric oscillators
\cite{marandi_network_2014,inagaki_large-scale_2016}, electromechanical
resonators \cite{Mahboob_mechIsing_2016}, and CMOS transistors
\cite{hayashi_accelerator_2016,Taiji_IsingProcessor_1988,Yamaoka_AIIsing_2016}.
Recently, individual exciton-polariton (polariton)
condensates~\cite{deng_condensation_2002,
    kasprzak_bose-einstein_2006,balili_bose-einstein_2007,
    baumberg_spontaneous_2008, schneider_electrically_2013,
    bhattacharya_solid_2013, daskalakis_nonlinear_2014,
    plumhof_room-temperature_2014} have been observed to spontaneously magnetize
\cite{ohadi_spontaneous_2015}, and when two condensates are close together the
spins can be controllably aligned (or anti-aligned)~\cite{ohadi_tunable_2016}.
Using these building blocks, we now explore the scaling up to a large 1D system,
constructing a non-equilibrium, driven-dissipative controlled spin-lattice of
exciton-polariton condensates. New types of order can appear in larger lattices,
while at the same time extra measures have to be taken to ensure scalability.

\begin{figure}
		\centering
		\includegraphics[width=1\linewidth]{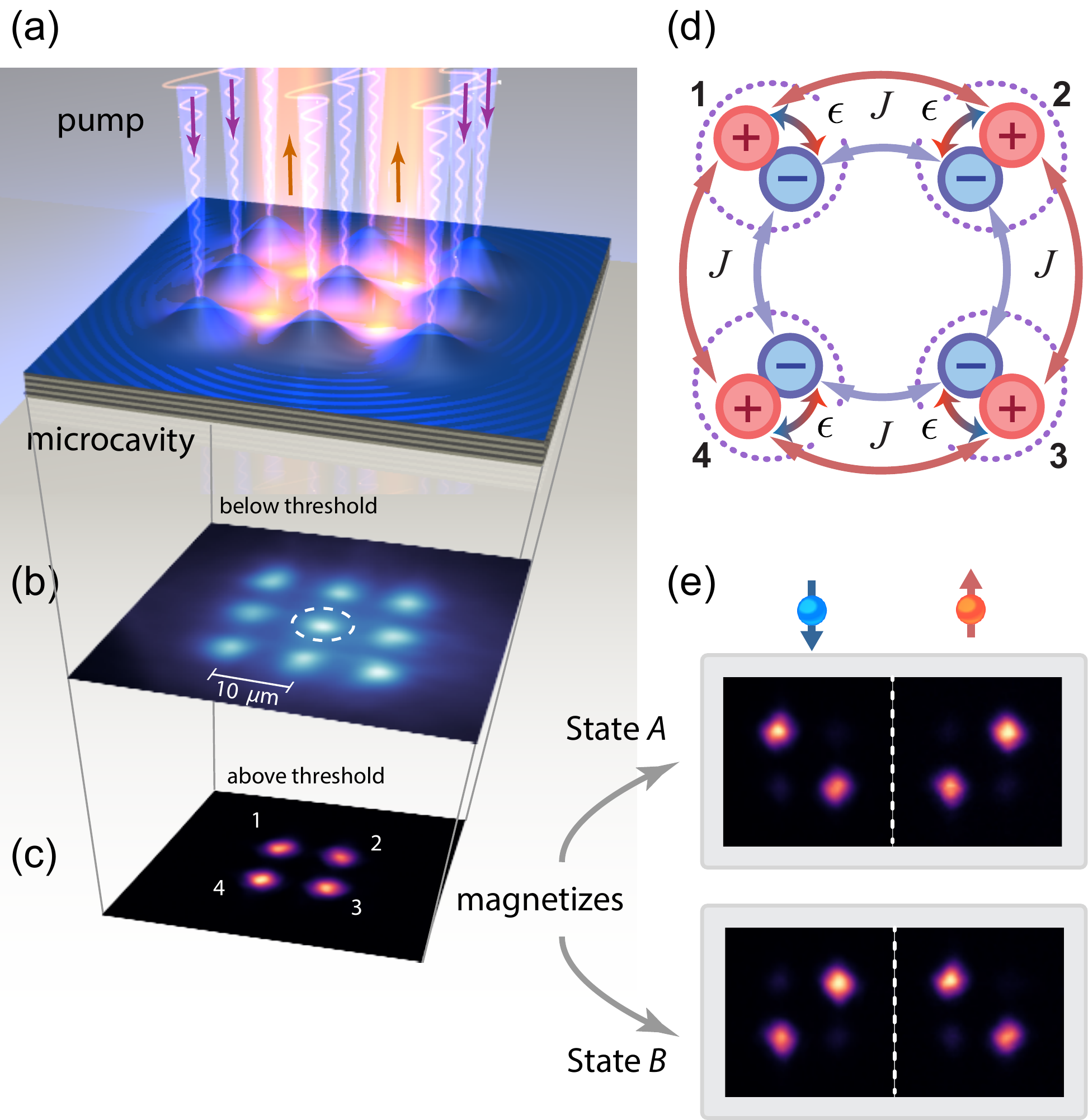}
		\caption{\capind{a}
		Lattice potential in the microcavity created by blueshifts at the
		pump beams (purple beams) forming magnetized condensates in
		the center of each site (yellow spots).
		\capind{b} Below threshold PL showing the pump spots. Global
		NN barrier is tuned by modulating the intensity of the center
		spot ($u_r$, dashed circle). \capind{c} Formation of condensates at the center of
each of the 4 lattice sites above threshold.
	    \capind{d} Schematic of condensate spin
		chain comprised of two coupled Bose-Hubbard chains. Each condensate
		(indices 1--4) has two spin states ($+$ and $-$)
		which are coherently coupled by $\epsilon$. Each is
		also coherently coupled to its same spin NN by Josephson coupling $J$.
		\capind{e} Magnetization of condensate chain (expt) above the spin-bifurcation
		threshold ($P_{\mathrm{c}}$), into an AFM state.
}\label{fig:1}
\end{figure}

Here, we study the spin properties of a closed interacting chain of
exciton-polariton condensates. When the pump laser is turned on, the system
spontaneously condenses into a magnetically ordered state on picosecond
timescales, and remains frozen in that state for many milliseconds. By optically
tuning the Josephson coupling between the condensates, the system can be tuned
from a ferromagnetic to anti-ferromagnetic phase, via a disordered crossover
phase. Remarkably, in a system of 4 identical spin condensates, where there is
no spatial disorder a paired spin state with alternating FM-AFM bonds is
observed.  Such a state cannot exist in a smaller system. Furthermore, despite
the larger phase degree-of-freedom offered by the larger spin chain, from
comparison to theory we conclude that the FM (AFM) bonds only adopt a
phase-shift of 0 ($\pi$) respectively. This locking of the phase and spin
effectively results in a binary spin system. As the system size is increased to
longer condensate chains, spatial inhomogeneity in the microcavity becomes an
issue. We demonstrate a strategy to engineer ferromagnetic, antiferromagnetic,
or glassy states of longer spin chains by simultaneously tailoring each
individual nearest neighbor (NN) coupling between sites. Our work introduces
interacting trapped polariton condensates as a controllable system for studying
complex non-linear spin models out of equilibrium.

Polaritons are mixed light-matter quasiparticles appearing due to the strong
coupling of photons in a microcavity and excitons in a semiconductor quantum
well~\cite{kavokin_microcavities_2007}. Polaritons are driven-dissipative bosons
which can condense into macroscopically coherent many-body
states~\cite{deng_condensation_2002, kasprzak_bose-einstein_2006,
balili_bose-einstein_2007, baumberg_spontaneous_2008}. High optical
accessibility, picosecond dynamics, large
non-linearity~\cite{savvidis_angle-resonant_2000} and other unique
properties~\cite{leyder_observation_2007, lagoudakis_observation_2009,
paraiso_multistability_2010, abbarchi_macroscopic_2013,
cristofolini_optical_2013, sala_spin-orbit_2015, dufferwiel_spin_2015}, with
potential application in semiconductor chip
devices~\cite{amo_exciton-polariton_2010, ballarini_all-optical_2013,
cerna_ultrafast_2013,
nguyen_realization_2013,schneider_electrically_2013,bhattacharya_solid_2013,
dreismann_sub-femtojoule_2016} make them particularly attractive.

Our system is a GaAs quantum-well microcavity (see SI.~\ref{si:sample}) with an optically-induced
two-dimensional square lattice potential where a magnetized polariton
condensate (emitting almost fully-circularly polarized light) forms at each
lattice site (Fig.~\ref{fig:1}a-c). We generate polaritons
by the non-resonant optical excitation of the microcavity. Each non-resonantly
pumped spot creates a local reservoir of hot excitons which rapidly lose energy
and flow out due to repulsion from hot excitons in the reservoir and repulsive
self-interactions. Therefore, the optical excitation acts as both the gain and
the trapping potential forming the lattice sites~\cite{wertz_spontaneous_2010,
tosi_sculpting_2012}. Polaritons scatter into the ground state by stimulated
scattering and final-state bosonic
amplification~\cite{savvidis_angle-resonant_2000}. Once the density at any site
surpasses a threshold, a macroscopically coherent condensate forms inside each
trap~\cite{cristofolini_optical_2013, askitopoulos_polariton_2013}.

\begin{figure}[tb]
	\centering
		\includegraphics[width=0.95\linewidth]{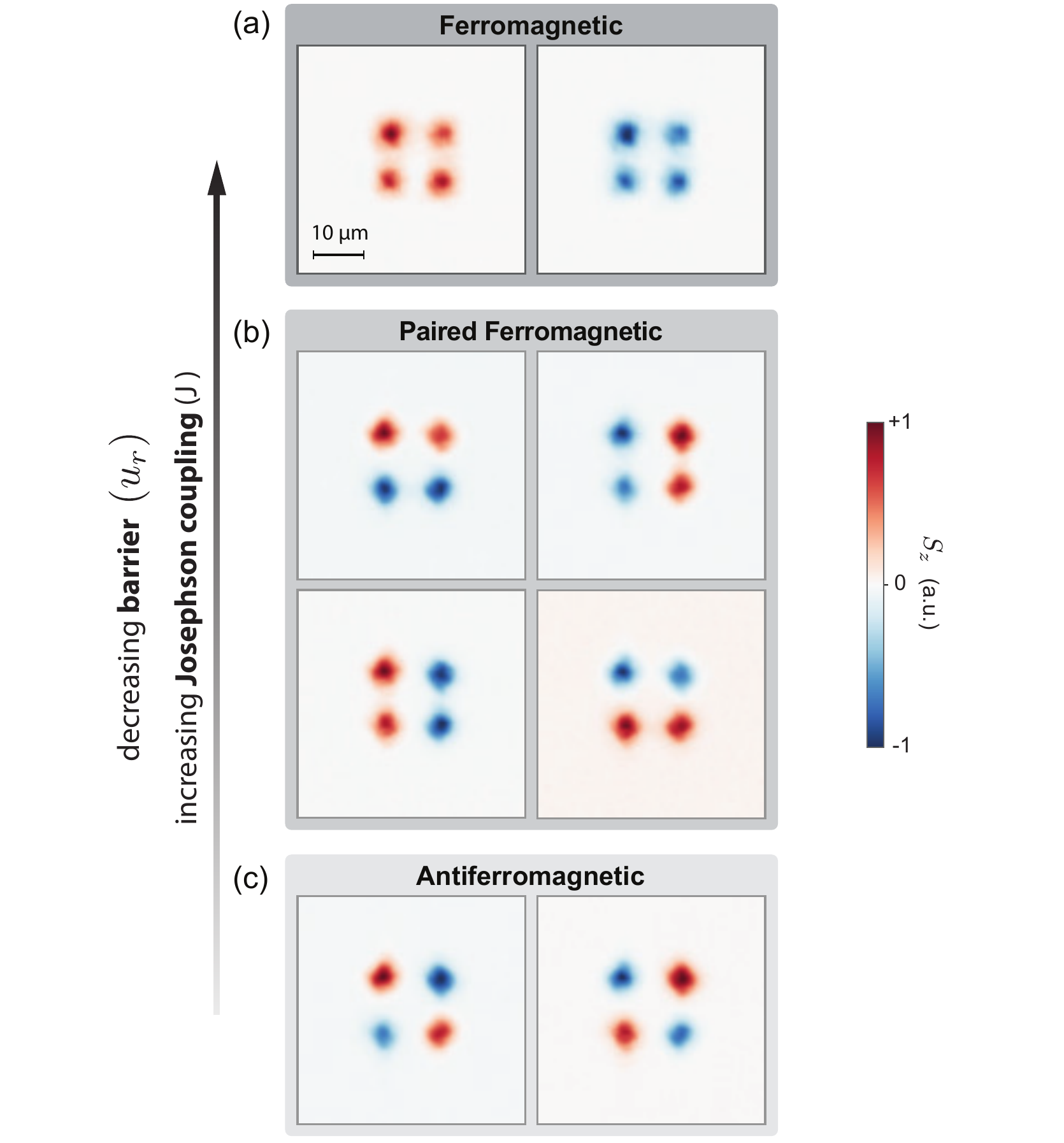}
		\caption{\captit{Steady states as a function of barrier height.} The
		measured condensate spin $S_z$ for all possible stable states at three
		phases of (a) FM, (b) PFM, and (c) AFM, when the global NN barrier
		$u_r$ is increased.}\label{fig:2}
\end{figure}

The total spin of the polaritons is quantized along the structure growth
axis, which corresponds to right- and left-circularly polarized photons emitted
from the cavity. For pump powers exceeding a spin-bifurcation threshold, trapped
polaritons can spontaneously magnetize by condensing into a single,
randomly-chosen spin state.  The spin-bifurcation process is driven by the
dissipation rate difference ($\gamma$) and energy splitting of the horizontally
and vertically polarized polaritons ($\epsilon$),  which determine the
spin-bifurcation threshold~\cite{ohadi_spontaneous_2015}.  We
operate above this threshold, which means each condensate spontaneously forms in
the spin-up or spin-down state with a degree of circular polarization
(condensate spin) $\vert S_z \vert > 85\%$.  Polariton condensates created by
the excitation pattern shown in Fig.~\ref{fig:1}b form a closed chain because
the potential from the central pump spot is so large that the tunneling of
polaritons between diagonal sites is negligible. Therefore, only the nearest
neighbor coupling is significant. In this geometry the spins of each condensate
are on-site coherently coupled by $\epsilon$, and each is coherently coupled to
that of the neighboring condensate by $J$ (Fig.~\ref{fig:1}d). By varying the
power of the central pump spot (Fig.~\ref{fig:1}b) we tune $J$, and by
changing the relative ratio of $J/\epsilon$ we can change the magnetic order of
the chain.  Remarkably, the condensate chains exhibit distinct magnetic phases
and mostly align their spins into particular patterns depending on
their coupling strengths. Initially we explore a 4-condensate system, before
showing how this behavior develops in the 8-condensate version.

\begin{figure*} \centering
	\includegraphics[width=1.0\textwidth]{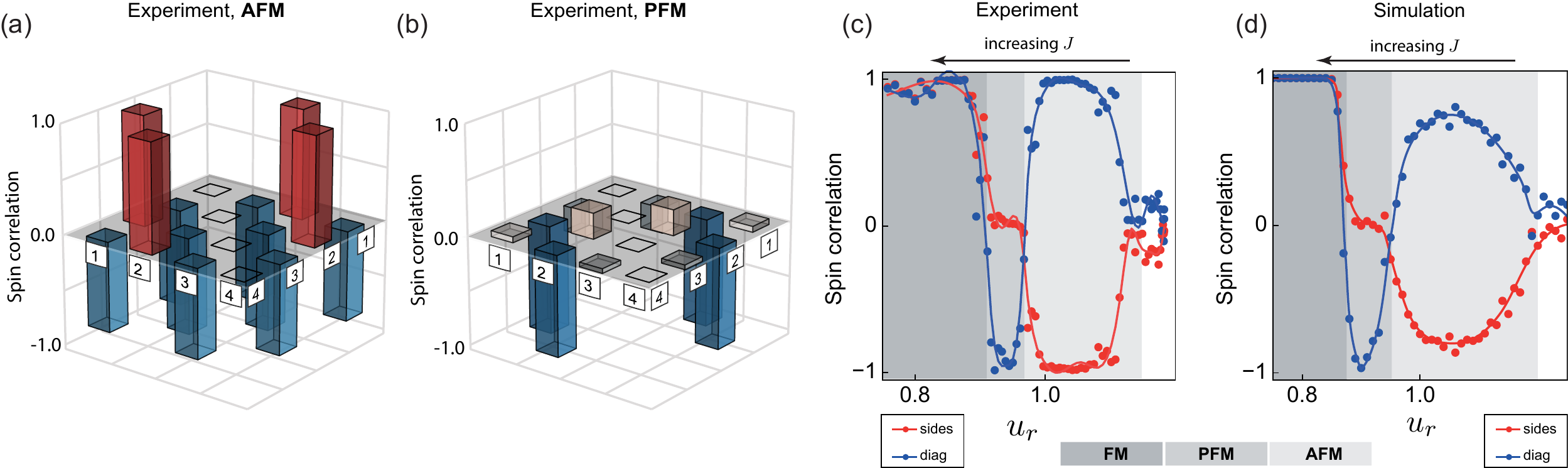}
	\caption{\captit{Spin correlations as a function of barrier height.}
	\capind{a,b} The measured correlation matrix of the spin chain at the two
	phases of (a) AFM and (b) PFM. Numbers show condensate indices. The auto-correlated diagonal elements are
	removed for clarity. \capind{c,d}  Phase diagram of the spin chain
	showing spin correlation of the diagonal and side condensates vs $u_r$
	in (c) experiment and (d) 2D numerical simulations.}\label{fig:3}
\end{figure*}


Magnetized condensation at the minima of the optically-induced lattice potential
is seen in the real-space photoluminescence (PL) from below to above the
condensation threshold (Fig.~\ref{fig:1}e). The critical magnetization
threshold is 1.3 times the condensation threshold.
We denote the intensity of the central pump spot relative to the rest of the
lattice spots by $u_r$. Since the local blueshift generated by the pump is
linearly proportional to the intensity of the pump spots, the barrier height
between each neighboring condensate increases as $u_r$ increases. Thus $u_r$ is
a measure of the coupling strength (Josephson tunneling rate $J$) of the condensate
lattice i.e. increasing $u_r$ corresponds to decreasing $J$. Since each condensate can form in the spin-up or the spin-down states
and couples with its two neighboring condensates, we expect to see the
formation of different spin patterns around the chain as we tune $u_r$.

We observe four distinct phases of the spin chain as we increase $u_r$
(Fig.~\ref{fig:2}): (1) FM with two spin degenerate states, formed from all
spin-up or spin-down states, (2) paired ferromagnetic (PFM) separated by two
domain walls, with four possible spin degenerate states and zero total spin, (3)
AFM with two possible spin degenerate states, and (4) paramagnetism with nearly
zero spin correlations between condensates (see also SI.~\ref{si:potlands}). In
each case, the spin chain spontaneously collapses into any of the degenerate
states due to random spin fluctuations from the reservoir at the onset of
magnetization. Although the final state of the chain is indeterminate for each
realization, once the spin chain forms it stays in that particular steady state
if a longer pump pulse (eg. \SI{100}{\milli\second}) is applied.

To characterize the spin correlations in each phase, we calculate the 4$\times$4
correlation matrix $C$ where the elements $C_{mn}=\rho(S_{z,m},S_{z,n})$ are the
Pearson correlation of spins of condensates $m$ and $n$ (as labelled
in Fig.~\ref{fig:1}c,d).  The correlation matrices for 100 realizations in the
AFM and PFM regimes (Fig.~\ref{fig:3}a,b) demonstrate robustly correlated spin
chains. To build their phase diagram we plot the average diagonal
$\bar{C}_{\mathrm{diag}}=(C_{13}+C_{24})/2$ and side
$\bar{C}_{\mathrm{side}}=(C_{12}+C_{23}+C_{34}+C_{41})/4$ condensate spin
correlations as a function of $u_r$ (Fig.~\ref{fig:3}c). We observe the FM phase
for $u_r<0.9$ followed by a sharp and narrow crossover to PFM for
$0.9<u_r<0.96$ and a second sharp crossover to a broad AFM phase for $0.96<u_r
<1.1$, succeeded by a rapid decay of correlations to near zero at higher $u_r$.
2D Ginzburg-Landau numerical simulations (see SI.~\ref{si:simulations})
accurately reproduce the experimental phase map (Fig.~\ref{fig:3}d).



\begin{figure*}[!tb] \centering \includegraphics[width=0.65\textwidth]{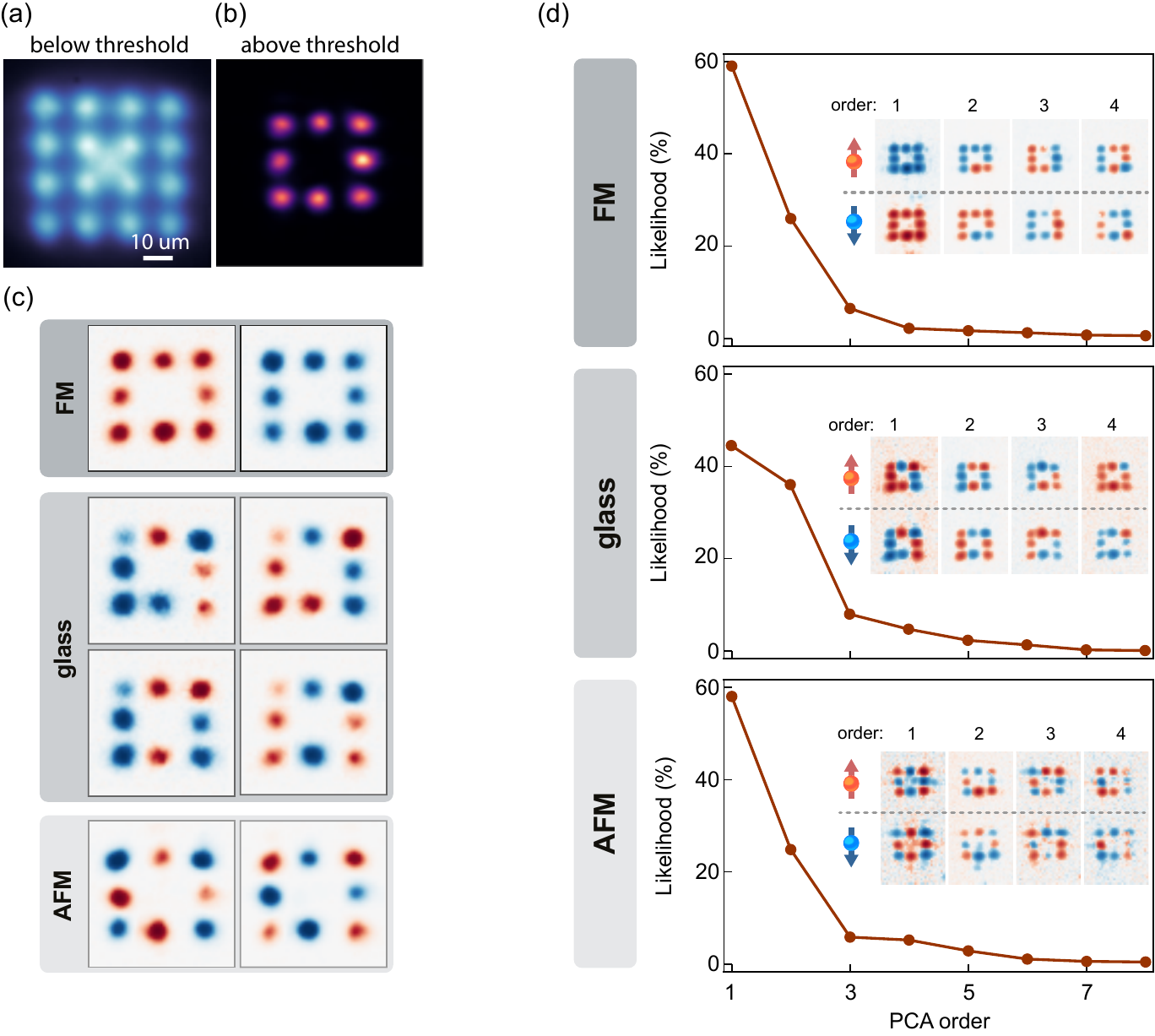}
    \caption{\captit{Controlling magnetic order with feedback}. \capind{a} PL
        below threshold showing excitation pattern. \capind{b} Emission above
        threshold showing 8-condensate chain.   \capind{c} The most probable
        spin states after optimization using FM, glass, or AFM search criteria
        in \textbf{d}. \capind{d} PCA components of spin-up/down
        emission after the search procedure for different
        targets. Intensities show strengths of
        components. Plots show likelihood (variance percentage) for each
        order.}\label{fig:4}
\end{figure*}


We can easily extend the square pumping geometry to accommodate longer spin
chains forming now a total of 8 condensates (Fig.~\ref{fig:4}a,b). Once again,
we observe FM, AFM, and a variety of spin glass states in this magnetic chain.
Because the number of barriers to modulate increases, their simultaneous control
is not as straightforward. At the same time, as the system size increases,
tiny spatial inhomogeneities in the microcavity become increasingly important.
The latter arise from the growth process and slightly change the local energy of
the polaritons, modulating the coupling strength between neighboring sites.  If
the energy modulation is large enough, it can even change the type of the
coupling at each bond.  Without more sophisticated approaches, this spatial
inhomogeneity of the microcavity would limit the size of condensate lattices
that can be studied, and thus prospects for using the system as a simulator.
This general issue is however generic in all condensate lattices.

We can, however, explore and correct for the spatial inhomogeneity here by tailoring
the imprinted excitation pattern. Since the background energy landscape is
unknown, we employ an iterative search algorithm with feedback to find the
optimal pattern needed to produce a desired correlated spin chain (see
SI.~\ref{si:feedback}).  At the end of each search process, which only takes a
few minutes, the most likely spin states can be inspected (Fig.~\ref{fig:4}c).
Principal component analysis (PCA) of the spin-up and spin-down intensities
reveals the most probable states after optimization (Fig.~\ref{fig:4}d).
In the FM and AFM phases, the pure states (1st PCA components)
are obtained in 60\% of instances, more than twice as likely as trapping
a single defect (2nd PCA component) with two domain walls. Other states
have $<$10\% probabilities.  By contrast, in the glass state we find
near-degenerate states with four domain walls that
dominate. Our 2-dimensional (2D) simulations show that a disorder potential of $\sim 5\mathrm{\mu eV}$  is enough
to break spin chain symmetry (see SI.~\ref{si:simulations}).   We thus show
this method can intialize the spin chain in any desired state.


We outline here a mean-field theory, extended from single trapped
condensates~\cite{aleiner_radiative_2012,ohadi_spontaneous_2015} to include the
Josephson coupling between nearest neighbors~\cite{ohadi_tunable_2016}. The
order parameter for each exciton-polariton condensate is a two-component
complex vector $\Psi_n=[\psi_{n+},\psi_{n-}]^\mathrm{T}$ where $\psi_{n+}$ and
$\psi_{n-}$ are the spin-up and spin-down wave functions of condensate $n$.
The order parameters evolve according to the driven dissipative equation:
\begin{align}
        \label{eq:main}
        i \dot{\Psi}_n = &-\frac{i}{2}g(S_n)\Psi_n-\frac{i}{2}(\gamma-i\epsilon)\sigma_x\Psi_n\\
        &+\frac{1}{2}(\bar{\alpha}S_n+\alpha S_{nz}\sigma_z)\Psi_n & \textrm{\footnotesize (interactions)}\nonumber\\
        &-\frac{J}{2}(\Psi_{n-1}+\Psi_{n+1}). & \textrm{\footnotesize (Josephson coupling)}\nonumber
\end{align}
Here, $g(S_n)=\Gamma-W+\eta S_n$ is the pumping-dissipation balance, $\Gamma$ is
the (average) dissipation rate, $W$ is the incoherent in-scattering,
$S_n=(\vert\psi_{n-}\vert^2+\vert\psi_{n+}\vert^2)/2$ and $\eta$ captures the
gain-saturation term~\cite{keeling_spontaneous_2008}. Linearly-polarized
single-polariton states in $X$ (horizontal) and $Y$ (vertical) are split in
energy by $\epsilon$ and in dissipation rate by $\gamma$, and $\sigma_{x,z}$ are
the Pauli matrices. The non-linear interaction constants are given by
$\bar{\alpha}=\alpha_1+\alpha_2$ and $\alpha=\alpha_1-\alpha_2$, where
$\alpha_1$ is the interaction constant for polaritons with the same spin and
$\alpha_2$ is the interaction constant for polaritons with opposite spins.
Finally, $J>0$ is the  spin-preserving Josephson
coupling~\cite{lagoudakis_coherent_2010,wouters_synchronized_2008,borgh_spatial_2010}
between nearest-neighbor condensates.

By making an ansatz where FM (AFM) bonds have a relative phase of 0 ($\pi$)
between nearest neighbors, we construct a mean-field model (see
SI.~\ref{si:qualitative}). This maps the system to a single condensate with an
energy shift $\omega_J$ and a renormalized polarization splitting $\epsilon_J$.
This allows us to apply the findings of ref. \cite{ohadi_spontaneous_2015} for a
single condensate to explain the phase-diagram of Fig.~\ref{fig:3}c,d, using two
criteria: (1) the final state must be stable, and (2) if multiple states are
stable, then the most probable final state is the one which turns magnetic at
the lowest power. To be stable requires $\epsilon_J>0$, so that on-site
spin coupling is strong enough to give magnetized condensates. In addition, the
spin-bifurcation threshold favors states with low $\epsilon_J$ (See Eq. S7).
The three most favorable spin phases then yield modified splittings:
$\epsilon_J^{\text{FM}}=\epsilon$, $\epsilon_J^{\text{glass}}=\epsilon-J$,
$\epsilon_J^{\text{AFM}}=\epsilon-2J$. Hence, the phase-diagram of
Fig.~\ref{fig:3}c is explained as follows. For $J<\epsilon/2$, all three states
are stable but the AFM state is favored since it has the lowest $\epsilon_J$.
For $\epsilon/2<J<\epsilon$, the AFM state becomes unstable and the glass state
is selected since it is now the lowest threshold state. For $J>\epsilon$, only
the FM state is stable. This explains all the key behaviours observed.

In conclusion, we demonstrate control of the spin states of closed chains of 4
and 8 polariton condensates. For small chains, the non-equilibrium
driven-dissipative spin lattice gives rise to a unique paired spin (paired-FM)
ordered state. This observation shows that our system is not governed
by the minimization of free-energy, as in for example the standard equilibrium Ising model. To
our knowledge, this paired-FM phase has not been observed in any equilibrium
or non-equilibrium binary spin system. In a 2D square lattice, in the paired-FM phase
each site must have two FM and two AFM bonds. Realizations of this phase can be mapped to different tilings of a chessboard with dominoes, which is a
\#P-complete problem~\cite{matousek_thirty-three_2010}.
We find that sample inhomogeneity hinders straightforward scaling to larger
chains. We overcome this problem
by careful feedback algorithms that compensate for sample inhomogeneities and
demonstrate a proof-of-principle scaling method. In the
absence of any corrections, the system behaves
 like spin glass, where interactions are randomly chosen by the
sample inhomogeneities acting as ``quenched disorder''.


\emph{Acknowledgments}---We acknowledge grants EPSRC EP/L027151/1, EU
INDEX 289968, ERC ``POLAFLOW'' Starting grant, ERC LINASS 320503, Spanish
MEC (MAT2008-01555), Mexican CONACYT 251808, EU FP7-REGPOT-2013-1 grant
agreement 316165 II, Leverhulme Trust Grant No. VP1-2013-011 and Fundación
La Caixa. H.S. and I.S. acknowledge support by the Research Fund of the
University of Iceland, The Icelandic Research Fund, Grant No. 163082-051.
TL was supported by the MOE AcRF Tier 1 grant 2016-T1-001-084.

\bibliography{bib}

\begin{thebibliography}{49}%
\makeatletter
\providecommand \@ifxundefined [1]{%
 \@ifx{#1\undefined}
}%
\providecommand \@ifnum [1]{%
 \ifnum #1\expandafter \@firstoftwo
 \else \expandafter \@secondoftwo
 \fi
}%
\providecommand \@ifx [1]{%
 \ifx #1\expandafter \@firstoftwo
 \else \expandafter \@secondoftwo
 \fi
}%
\providecommand \natexlab [1]{#1}%
\providecommand \enquote  [1]{``#1''}%
\providecommand \bibnamefont  [1]{#1}%
\providecommand \bibfnamefont [1]{#1}%
\providecommand \citenamefont [1]{#1}%
\providecommand \href@noop [0]{\@secondoftwo}%
\providecommand \href [0]{\begingroup \@sanitize@url \@href}%
\providecommand \@href[1]{\@@startlink{#1}\@@href}%
\providecommand \@@href[1]{\endgroup#1\@@endlink}%
\providecommand \@sanitize@url [0]{\catcode `\\12\catcode `\$12\catcode
  `\&12\catcode `\#12\catcode `\^12\catcode `\_12\catcode `\%12\relax}%
\providecommand \@@startlink[1]{}%
\providecommand \@@endlink[0]{}%
\providecommand \url  [0]{\begingroup\@sanitize@url \@url }%
\providecommand \@url [1]{\endgroup\@href {#1}{\urlprefix }}%
\providecommand \urlprefix  [0]{URL }%
\providecommand \Eprint [0]{\href }%
\providecommand \doibase [0]{http://dx.doi.org/}%
\providecommand \selectlanguage [0]{\@gobble}%
\providecommand \bibinfo  [0]{\@secondoftwo}%
\providecommand \bibfield  [0]{\@secondoftwo}%
\providecommand \translation [1]{[#1]}%
\providecommand \BibitemOpen [0]{}%
\providecommand \bibitemStop [0]{}%
\providecommand \bibitemNoStop [0]{.\EOS\space}%
\providecommand \EOS [0]{\spacefactor3000\relax}%
\providecommand \BibitemShut  [1]{\csname bibitem#1\endcsname}%
\let\auto@bib@innerbib\@empty
\bibitem [{\citenamefont {Baxter}(1989)}]{baxter_exactly_2016}%
  \BibitemOpen
  \bibfield  {author} {\bibinfo {author} {\bibfnamefont {R.~J.}\ \bibnamefont
  {Baxter}},\ }\href@noop {} {\emph {\bibinfo {title} {Exactly Solved Models in
  Statistical Mechanics}}}\ (\bibinfo  {publisher} {Elsevier},\ \bibinfo {year}
  {1989})\BibitemShut {NoStop}%
\bibitem [{\citenamefont {Kirkpatrick}\ \emph {et~al.}(1983)\citenamefont
  {Kirkpatrick}, \citenamefont {Gelatt},\ and\ \citenamefont
  {Vecchi}}]{kirkpatrick_optimization_1983}%
  \BibitemOpen
  \bibfield  {author} {\bibinfo {author} {\bibfnamefont {S.}~\bibnamefont
  {Kirkpatrick}}, \bibinfo {author} {\bibfnamefont {C.~D.}\ \bibnamefont
  {Gelatt}}, \ and\ \bibinfo {author} {\bibfnamefont {M.~P.}\ \bibnamefont
  {Vecchi}},\ }\href {\doibase 10.1126/science.220.4598.671} {\bibfield
  {journal} {\bibinfo  {journal} {Science}\ }\textbf {\bibinfo {volume}
  {220}},\ \bibinfo {pages} {671} (\bibinfo {year} {1983})}\BibitemShut
  {NoStop}%
\bibitem [{\citenamefont {Lucas}(2014)}]{lucas_ising_2014}%
  \BibitemOpen
  \bibfield  {author} {\bibinfo {author} {\bibfnamefont {A.}~\bibnamefont
  {Lucas}},\ }\href {\doibase 10.3389/fphy.2014.00005} {\bibfield  {journal}
  {\bibinfo  {journal} {Front. Physics}\ }\textbf {\bibinfo {volume} {2}},\
  \bibinfo {pages} {5} (\bibinfo {year} {2014})}\BibitemShut {NoStop}%
\bibitem [{\citenamefont {Hayashi}\ \emph {et~al.}(2016)\citenamefont
  {Hayashi}, \citenamefont {Yamaoka}, \citenamefont {Yoshimura}, \citenamefont
  {Okuyama}, \citenamefont {Aoki},\ and\ \citenamefont
  {Mizuno}}]{hayashi_accelerator_2016}%
  \BibitemOpen
  \bibfield  {author} {\bibinfo {author} {\bibfnamefont {M.}~\bibnamefont
  {Hayashi}}, \bibinfo {author} {\bibfnamefont {M.}~\bibnamefont {Yamaoka}},
  \bibinfo {author} {\bibfnamefont {C.}~\bibnamefont {Yoshimura}}, \bibinfo
  {author} {\bibfnamefont {T.}~\bibnamefont {Okuyama}}, \bibinfo {author}
  {\bibfnamefont {H.}~\bibnamefont {Aoki}}, \ and\ \bibinfo {author}
  {\bibfnamefont {H.}~\bibnamefont {Mizuno}},\ }\href
  {http://www.ijnc.org/index.php/ijnc/article/view/125} {\bibfield  {journal}
  {\bibinfo  {journal} {Int. J. Net. Comp.}\ }\textbf {\bibinfo {volume} {6}},\
  \bibinfo {pages} {195} (\bibinfo {year} {2016})}\BibitemShut {NoStop}%
\bibitem [{\citenamefont {Yamaoka}\ \emph {et~al.}(2016)\citenamefont
  {Yamaoka}, \citenamefont {Yoshimura}, \citenamefont {Hayashi}, \citenamefont
  {Okuyama}, \citenamefont {Aoki},\ and\ \citenamefont
  {Mizuno}}]{Yamaoka_AIIsing_2016}%
  \BibitemOpen
  \bibfield  {author} {\bibinfo {author} {\bibfnamefont {M.}~\bibnamefont
  {Yamaoka}}, \bibinfo {author} {\bibfnamefont {C.}~\bibnamefont {Yoshimura}},
  \bibinfo {author} {\bibfnamefont {M.}~\bibnamefont {Hayashi}}, \bibinfo
  {author} {\bibfnamefont {T.}~\bibnamefont {Okuyama}}, \bibinfo {author}
  {\bibfnamefont {H.}~\bibnamefont {Aoki}}, \ and\ \bibinfo {author}
  {\bibfnamefont {H.}~\bibnamefont {Mizuno}},\ }\href
  {http://www.hitachi.com/rev/pdf/2016/r2016_06_110.pdf} {\bibfield  {journal}
  {\bibinfo  {journal} {Hitachi Rev.}\ }\textbf {\bibinfo {volume} {65}},\
  \bibinfo {pages} {156} (\bibinfo {year} {2016})}\BibitemShut {NoStop}%
\bibitem [{\citenamefont {Barahona}(1982)}]{barahona_computational_1982}%
  \BibitemOpen
  \bibfield  {author} {\bibinfo {author} {\bibfnamefont {F.}~\bibnamefont
  {Barahona}},\ }\href {http://stacks.iop.org/0305-4470/15/i=10/a=028}
  {\bibfield  {journal} {\bibinfo  {journal} {J. Phys. A: Math. Gen.}\ }\textbf
  {\bibinfo {volume} {15}},\ \bibinfo {pages} {3241} (\bibinfo {year}
  {1982})}\BibitemShut {NoStop}%
\bibitem [{\citenamefont {Buluta}\ and\ \citenamefont
  {Nori}(2009)}]{buluta_quantum_2009}%
  \BibitemOpen
  \bibfield  {author} {\bibinfo {author} {\bibfnamefont {I.}~\bibnamefont
  {Buluta}}\ and\ \bibinfo {author} {\bibfnamefont {F.}~\bibnamefont {Nori}},\
  }\href {\doibase 10.1126/science.1177838} {\bibfield  {journal} {\bibinfo
  {journal} {Science}\ }\textbf {\bibinfo {volume} {326}},\ \bibinfo {pages}
  {108} (\bibinfo {year} {2009})}\BibitemShut {NoStop}%
\bibitem [{\citenamefont {Georgescu}\ \emph {et~al.}(2014)\citenamefont
  {Georgescu}, \citenamefont {Ashhab},\ and\ \citenamefont
  {Nori}}]{georgescu_quantum_2014}%
  \BibitemOpen
  \bibfield  {author} {\bibinfo {author} {\bibfnamefont {I.~M.}\ \bibnamefont
  {Georgescu}}, \bibinfo {author} {\bibfnamefont {S.}~\bibnamefont {Ashhab}}, \
  and\ \bibinfo {author} {\bibfnamefont {F.}~\bibnamefont {Nori}},\ }\href
  {\doibase 10.1103/RevModPhys.86.153} {\bibfield  {journal} {\bibinfo
  {journal} {Rev. Mod. Phys.}\ }\textbf {\bibinfo {volume} {86}},\ \bibinfo
  {pages} {153} (\bibinfo {year} {2014})}\BibitemShut {NoStop}%
\bibitem [{\citenamefont {Bloch}\ \emph {et~al.}(2012)\citenamefont {Bloch},
  \citenamefont {Dalibard},\ and\ \citenamefont
  {Nascimbène}}]{bloch_quantum_2012}%
  \BibitemOpen
  \bibfield  {author} {\bibinfo {author} {\bibfnamefont {I.}~\bibnamefont
  {Bloch}}, \bibinfo {author} {\bibfnamefont {J.}~\bibnamefont {Dalibard}}, \
  and\ \bibinfo {author} {\bibfnamefont {S.}~\bibnamefont {Nascimbène}},\
  }\href {\doibase 10.1038/nphys2259} {\bibfield  {journal} {\bibinfo
  {journal} {Nature Phys.}\ }\textbf {\bibinfo {volume} {8}},\ \bibinfo {pages}
  {267} (\bibinfo {year} {2012})}\BibitemShut {NoStop}%
\bibitem [{\citenamefont {Marandi}\ \emph {et~al.}(2014)\citenamefont
  {Marandi}, \citenamefont {Wang}, \citenamefont {Takata}, \citenamefont
  {Byer},\ and\ \citenamefont {Yamamoto}}]{marandi_network_2014}%
  \BibitemOpen
  \bibfield  {author} {\bibinfo {author} {\bibfnamefont {A.}~\bibnamefont
  {Marandi}}, \bibinfo {author} {\bibfnamefont {Z.}~\bibnamefont {Wang}},
  \bibinfo {author} {\bibfnamefont {K.}~\bibnamefont {Takata}}, \bibinfo
  {author} {\bibfnamefont {R.~L.}\ \bibnamefont {Byer}}, \ and\ \bibinfo
  {author} {\bibfnamefont {Y.}~\bibnamefont {Yamamoto}},\ }\href {\doibase
  10.1038/nphoton.2014.249} {\bibfield  {journal} {\bibinfo  {journal} {Nature
  Photon.}\ }\textbf {\bibinfo {volume} {8}},\ \bibinfo {pages} {937} (\bibinfo
  {year} {2014})}\BibitemShut {NoStop}%
\bibitem [{\citenamefont {Inagaki}\ \emph {et~al.}(2016)\citenamefont
  {Inagaki}, \citenamefont {Inaba}, \citenamefont {Hamerly}, \citenamefont
  {Inoue}, \citenamefont {Yamamoto},\ and\ \citenamefont
  {Takesue}}]{inagaki_large-scale_2016}%
  \BibitemOpen
  \bibfield  {author} {\bibinfo {author} {\bibfnamefont {T.}~\bibnamefont
  {Inagaki}}, \bibinfo {author} {\bibfnamefont {K.}~\bibnamefont {Inaba}},
  \bibinfo {author} {\bibfnamefont {R.}~\bibnamefont {Hamerly}}, \bibinfo
  {author} {\bibfnamefont {K.}~\bibnamefont {Inoue}}, \bibinfo {author}
  {\bibfnamefont {Y.}~\bibnamefont {Yamamoto}}, \ and\ \bibinfo {author}
  {\bibfnamefont {H.}~\bibnamefont {Takesue}},\ }\href {\doibase
  10.1038/nphoton.2016.68} {\bibfield  {journal} {\bibinfo  {journal} {Nature
  Photon.}\ }\textbf {\bibinfo {volume} {10}},\ \bibinfo {pages} {415}
  (\bibinfo {year} {2016})}\BibitemShut {NoStop}%
\bibitem [{\citenamefont {Mahboob}\ \emph {et~al.}(2016)\citenamefont
  {Mahboob}, \citenamefont {Okamoto},\ and\ \citenamefont
  {Yamaguchi}}]{Mahboob_mechIsing_2016}%
  \BibitemOpen
  \bibfield  {author} {\bibinfo {author} {\bibfnamefont {I.}~\bibnamefont
  {Mahboob}}, \bibinfo {author} {\bibfnamefont {H.}~\bibnamefont {Okamoto}}, \
  and\ \bibinfo {author} {\bibfnamefont {H.}~\bibnamefont {Yamaguchi}},\ }\href
  {\doibase 10.1126/sciadv.1600236} {\bibfield  {journal} {\bibinfo  {journal}
  {Sci. Adv.}\ }\textbf {\bibinfo {volume} {2}},\ \bibinfo {pages} {e1600236}
  (\bibinfo {year} {2016})}\BibitemShut {NoStop}%
\bibitem [{\citenamefont {Taiji}\ \emph {et~al.}(1988)\citenamefont {Taiji},
  \citenamefont {Ito},\ and\ \citenamefont
  {Suzuki}}]{Taiji_IsingProcessor_1988}%
  \BibitemOpen
  \bibfield  {author} {\bibinfo {author} {\bibfnamefont {M.}~\bibnamefont
  {Taiji}}, \bibinfo {author} {\bibfnamefont {N.}~\bibnamefont {Ito}}, \ and\
  \bibinfo {author} {\bibfnamefont {M.}~\bibnamefont {Suzuki}},\ }\href
  {\doibase doi: 10.1063/1.1139934} {\bibfield  {journal} {\bibinfo  {journal}
  {Rev. Sci. Instrument.}\ }\textbf {\bibinfo {volume} {59}},\ \bibinfo {pages}
  {2483} (\bibinfo {year} {1988})}\BibitemShut {NoStop}%
\bibitem [{\citenamefont {Deng}\ \emph {et~al.}(2002)\citenamefont {Deng},
  \citenamefont {Weihs}, \citenamefont {Santori}, \citenamefont {Bloch},\ and\
  \citenamefont {Yamamoto}}]{deng_condensation_2002}%
  \BibitemOpen
  \bibfield  {author} {\bibinfo {author} {\bibfnamefont {H.}~\bibnamefont
  {Deng}}, \bibinfo {author} {\bibfnamefont {G.}~\bibnamefont {Weihs}},
  \bibinfo {author} {\bibfnamefont {C.}~\bibnamefont {Santori}}, \bibinfo
  {author} {\bibfnamefont {J.}~\bibnamefont {Bloch}}, \ and\ \bibinfo {author}
  {\bibfnamefont {Y.}~\bibnamefont {Yamamoto}},\ }\href {\doibase
  10.1126/science.1074464} {\bibfield  {journal} {\bibinfo  {journal}
  {Science}\ }\textbf {\bibinfo {volume} {298}},\ \bibinfo {pages} {199}
  (\bibinfo {year} {2002})}\BibitemShut {NoStop}%
\bibitem [{\citenamefont {Kasprzak}\ \emph {et~al.}(2006)\citenamefont
  {Kasprzak}, \citenamefont {Richard}, \citenamefont {Kundermann},
  \citenamefont {Baas}, \citenamefont {Jeambrun}, \citenamefont {Keeling},
  \citenamefont {Marchetti}, \citenamefont {Szymańska}, \citenamefont
  {André}, \citenamefont {Staehli} \emph
  {et~al.}}]{kasprzak_bose-einstein_2006}%
  \BibitemOpen
  \bibfield  {author} {\bibinfo {author} {\bibfnamefont {J.}~\bibnamefont
  {Kasprzak}}, \bibinfo {author} {\bibfnamefont {M.}~\bibnamefont {Richard}},
  \bibinfo {author} {\bibfnamefont {S.}~\bibnamefont {Kundermann}}, \bibinfo
  {author} {\bibfnamefont {A.}~\bibnamefont {Baas}}, \bibinfo {author}
  {\bibfnamefont {P.}~\bibnamefont {Jeambrun}}, \bibinfo {author}
  {\bibfnamefont {J.~M.~J.}\ \bibnamefont {Keeling}}, \bibinfo {author}
  {\bibfnamefont {F.~M.}\ \bibnamefont {Marchetti}}, \bibinfo {author}
  {\bibfnamefont {M.~H.}\ \bibnamefont {Szymańska}}, \bibinfo {author}
  {\bibfnamefont {R.}~\bibnamefont {André}}, \bibinfo {author} {\bibfnamefont
  {J.~L.}\ \bibnamefont {Staehli}},  \emph {et~al.},\ }\href {\doibase
  10.1038/nature05131} {\bibfield  {journal} {\bibinfo  {journal} {Nature}\
  }\textbf {\bibinfo {volume} {443}},\ \bibinfo {pages} {409} (\bibinfo {year}
  {2006})}\BibitemShut {NoStop}%
\bibitem [{\citenamefont {Balili}\ \emph {et~al.}(2007)\citenamefont {Balili},
  \citenamefont {Hartwell}, \citenamefont {Snoke}, \citenamefont {Pfeiffer},\
  and\ \citenamefont {West}}]{balili_bose-einstein_2007}%
  \BibitemOpen
  \bibfield  {author} {\bibinfo {author} {\bibfnamefont {R.}~\bibnamefont
  {Balili}}, \bibinfo {author} {\bibfnamefont {V.}~\bibnamefont {Hartwell}},
  \bibinfo {author} {\bibfnamefont {D.}~\bibnamefont {Snoke}}, \bibinfo
  {author} {\bibfnamefont {L.}~\bibnamefont {Pfeiffer}}, \ and\ \bibinfo
  {author} {\bibfnamefont {K.}~\bibnamefont {West}},\ }\href {\doibase
  10.1126/science.1140990} {\bibfield  {journal} {\bibinfo  {journal}
  {Science}\ }\textbf {\bibinfo {volume} {316}},\ \bibinfo {pages} {1007}
  (\bibinfo {year} {2007})}\BibitemShut {NoStop}%
\bibitem [{\citenamefont {Baumberg}\ \emph {et~al.}(2008)\citenamefont
  {Baumberg}, \citenamefont {Kavokin}, \citenamefont {Christopoulos},
  \citenamefont {Grundy}, \citenamefont {Butté}, \citenamefont {Christmann},
  \citenamefont {Solnyshkov}, \citenamefont {Malpuech}, \citenamefont
  {Baldassarri Höger~von Högersthal}, \citenamefont {Feltin} \emph
  {et~al.}}]{baumberg_spontaneous_2008}%
  \BibitemOpen
  \bibfield  {author} {\bibinfo {author} {\bibfnamefont {J.}~\bibnamefont
  {Baumberg}}, \bibinfo {author} {\bibfnamefont {A.}~\bibnamefont {Kavokin}},
  \bibinfo {author} {\bibfnamefont {S.}~\bibnamefont {Christopoulos}}, \bibinfo
  {author} {\bibfnamefont {A.}~\bibnamefont {Grundy}}, \bibinfo {author}
  {\bibfnamefont {R.}~\bibnamefont {Butté}}, \bibinfo {author} {\bibfnamefont
  {G.}~\bibnamefont {Christmann}}, \bibinfo {author} {\bibfnamefont
  {D.}~\bibnamefont {Solnyshkov}}, \bibinfo {author} {\bibfnamefont
  {G.}~\bibnamefont {Malpuech}}, \bibinfo {author} {\bibfnamefont
  {G.}~\bibnamefont {Baldassarri Höger~von Högersthal}}, \bibinfo {author}
  {\bibfnamefont {E.}~\bibnamefont {Feltin}},  \emph {et~al.},\ }\href
  {\doibase 10.1103/PhysRevLett.101.136409} {\bibfield  {journal} {\bibinfo
  {journal} {Phys. Rev. Lett.}\ }\textbf {\bibinfo {volume} {101}},\ \bibinfo
  {pages} {136409} (\bibinfo {year} {2008})}\BibitemShut {NoStop}%
\bibitem [{\citenamefont {Schneider}\ \emph {et~al.}(2013)\citenamefont
  {Schneider}, \citenamefont {Rahimi-Iman}, \citenamefont {Kim}, \citenamefont
  {Fischer}, \citenamefont {Savenko}, \citenamefont {Amthor}, \citenamefont
  {Lermer}, \citenamefont {Wolf}, \citenamefont {Worschech}, \citenamefont
  {Kulakovskii}, \citenamefont {Shelykh}, \citenamefont {Kamp}, \citenamefont
  {Reitzenstein}, \citenamefont {Forchel}, \citenamefont {Yamamoto},\ and\
  \citenamefont {Höfling}}]{schneider_electrically_2013}%
  \BibitemOpen
  \bibfield  {author} {\bibinfo {author} {\bibfnamefont {C.}~\bibnamefont
  {Schneider}}, \bibinfo {author} {\bibfnamefont {A.}~\bibnamefont
  {Rahimi-Iman}}, \bibinfo {author} {\bibfnamefont {N.~Y.}\ \bibnamefont
  {Kim}}, \bibinfo {author} {\bibfnamefont {J.}~\bibnamefont {Fischer}},
  \bibinfo {author} {\bibfnamefont {I.~G.}\ \bibnamefont {Savenko}}, \bibinfo
  {author} {\bibfnamefont {M.}~\bibnamefont {Amthor}}, \bibinfo {author}
  {\bibfnamefont {M.}~\bibnamefont {Lermer}}, \bibinfo {author} {\bibfnamefont
  {A.}~\bibnamefont {Wolf}}, \bibinfo {author} {\bibfnamefont {L.}~\bibnamefont
  {Worschech}}, \bibinfo {author} {\bibfnamefont {V.~D.}\ \bibnamefont
  {Kulakovskii}}, \bibinfo {author} {\bibfnamefont {I.~A.}\ \bibnamefont
  {Shelykh}}, \bibinfo {author} {\bibfnamefont {M.}~\bibnamefont {Kamp}},
  \bibinfo {author} {\bibfnamefont {S.}~\bibnamefont {Reitzenstein}}, \bibinfo
  {author} {\bibfnamefont {A.}~\bibnamefont {Forchel}}, \bibinfo {author}
  {\bibfnamefont {Y.}~\bibnamefont {Yamamoto}}, \ and\ \bibinfo {author}
  {\bibfnamefont {S.}~\bibnamefont {Höfling}},\ }\href {\doibase
  10.1038/nature12036} {\bibfield  {journal} {\bibinfo  {journal} {Nature}\
  }\textbf {\bibinfo {volume} {497}},\ \bibinfo {pages} {348} (\bibinfo {year}
  {2013})}\BibitemShut {NoStop}%
\bibitem [{\citenamefont {Bhattacharya}\ \emph {et~al.}(2013)\citenamefont
  {Bhattacharya}, \citenamefont {Xiao}, \citenamefont {Das}, \citenamefont
  {Bhowmick},\ and\ \citenamefont {Heo}}]{bhattacharya_solid_2013}%
  \BibitemOpen
  \bibfield  {author} {\bibinfo {author} {\bibfnamefont {P.}~\bibnamefont
  {Bhattacharya}}, \bibinfo {author} {\bibfnamefont {B.}~\bibnamefont {Xiao}},
  \bibinfo {author} {\bibfnamefont {A.}~\bibnamefont {Das}}, \bibinfo {author}
  {\bibfnamefont {S.}~\bibnamefont {Bhowmick}}, \ and\ \bibinfo {author}
  {\bibfnamefont {J.}~\bibnamefont {Heo}},\ }\href {\doibase
  10.1103/PhysRevLett.110.206403} {\bibfield  {journal} {\bibinfo  {journal}
  {Phys. Rev. Lett.}\ }\textbf {\bibinfo {volume} {110}},\ \bibinfo {pages}
  {206403} (\bibinfo {year} {2013})}\BibitemShut {NoStop}%
\bibitem [{\citenamefont {Daskalakis}\ \emph {et~al.}(2014)\citenamefont
  {Daskalakis}, \citenamefont {Maier}, \citenamefont {Murray},\ and\
  \citenamefont {Kéna-Cohen}}]{daskalakis_nonlinear_2014}%
  \BibitemOpen
  \bibfield  {author} {\bibinfo {author} {\bibfnamefont {K.~S.}\ \bibnamefont
  {Daskalakis}}, \bibinfo {author} {\bibfnamefont {S.~A.}\ \bibnamefont
  {Maier}}, \bibinfo {author} {\bibfnamefont {R.}~\bibnamefont {Murray}}, \
  and\ \bibinfo {author} {\bibfnamefont {S.}~\bibnamefont {Kéna-Cohen}},\
  }\href {\doibase 10.1038/nmat3874} {\bibfield  {journal} {\bibinfo  {journal}
  {Nat. Mater.}\ }\textbf {\bibinfo {volume} {13}},\ \bibinfo {pages} {271}
  (\bibinfo {year} {2014})}\BibitemShut {NoStop}%
\bibitem [{\citenamefont {Plumhof}\ \emph {et~al.}(2014)\citenamefont
  {Plumhof}, \citenamefont {Stöferle}, \citenamefont {Mai}, \citenamefont
  {Scherf},\ and\ \citenamefont {Mahrt}}]{plumhof_room-temperature_2014}%
  \BibitemOpen
  \bibfield  {author} {\bibinfo {author} {\bibfnamefont {J.~D.}\ \bibnamefont
  {Plumhof}}, \bibinfo {author} {\bibfnamefont {T.}~\bibnamefont {Stöferle}},
  \bibinfo {author} {\bibfnamefont {L.}~\bibnamefont {Mai}}, \bibinfo {author}
  {\bibfnamefont {U.}~\bibnamefont {Scherf}}, \ and\ \bibinfo {author}
  {\bibfnamefont {R.~F.}\ \bibnamefont {Mahrt}},\ }\href {\doibase
  10.1038/nmat3825} {\bibfield  {journal} {\bibinfo  {journal} {Nat. Mater.}\
  }\textbf {\bibinfo {volume} {13}},\ \bibinfo {pages} {247} (\bibinfo {year}
  {2014})}\BibitemShut {NoStop}%
\bibitem [{\citenamefont {Ohadi}\ \emph {et~al.}(2015)\citenamefont {Ohadi},
  \citenamefont {Dreismann}, \citenamefont {Rubo}, \citenamefont {Pinsker},
  \citenamefont {del Valle-Inclan~Redondo}, \citenamefont {Tsintzos},
  \citenamefont {Hatzopoulos}, \citenamefont {Savvidis},\ and\ \citenamefont
  {Baumberg}}]{ohadi_spontaneous_2015}%
  \BibitemOpen
  \bibfield  {author} {\bibinfo {author} {\bibfnamefont {H.}~\bibnamefont
  {Ohadi}}, \bibinfo {author} {\bibfnamefont {A.}~\bibnamefont {Dreismann}},
  \bibinfo {author} {\bibfnamefont {Y.~G.}\ \bibnamefont {Rubo}}, \bibinfo
  {author} {\bibfnamefont {F.}~\bibnamefont {Pinsker}}, \bibinfo {author}
  {\bibfnamefont {Y.}~\bibnamefont {del Valle-Inclan~Redondo}}, \bibinfo
  {author} {\bibfnamefont {S.~I.}\ \bibnamefont {Tsintzos}}, \bibinfo {author}
  {\bibfnamefont {Z.}~\bibnamefont {Hatzopoulos}}, \bibinfo {author}
  {\bibfnamefont {P.~G.}\ \bibnamefont {Savvidis}}, \ and\ \bibinfo {author}
  {\bibfnamefont {J.~J.}\ \bibnamefont {Baumberg}},\ }\href {\doibase
  10.1103/PhysRevX.5.031002} {\bibfield  {journal} {\bibinfo  {journal} {Phys.
  Rev. X}\ }\textbf {\bibinfo {volume} {5}},\ \bibinfo {pages} {031002}
  (\bibinfo {year} {2015})}\BibitemShut {NoStop}%
\bibitem [{\citenamefont {Ohadi}\ \emph {et~al.}(2016)\citenamefont {Ohadi},
  \citenamefont {del Valle-Inclan~Redondo}, \citenamefont {Dreismann},
  \citenamefont {Rubo}, \citenamefont {Pinsker}, \citenamefont {Tsintzos},
  \citenamefont {Hatzopoulos}, \citenamefont {Savvidis},\ and\ \citenamefont
  {Baumberg}}]{ohadi_tunable_2016}%
  \BibitemOpen
  \bibfield  {author} {\bibinfo {author} {\bibfnamefont {H.}~\bibnamefont
  {Ohadi}}, \bibinfo {author} {\bibfnamefont {Y.}~\bibnamefont {del
  Valle-Inclan~Redondo}}, \bibinfo {author} {\bibfnamefont {A.}~\bibnamefont
  {Dreismann}}, \bibinfo {author} {\bibfnamefont {Y.~G.}\ \bibnamefont {Rubo}},
  \bibinfo {author} {\bibfnamefont {F.}~\bibnamefont {Pinsker}}, \bibinfo
  {author} {\bibfnamefont {S.~I.}\ \bibnamefont {Tsintzos}}, \bibinfo {author}
  {\bibfnamefont {Z.}~\bibnamefont {Hatzopoulos}}, \bibinfo {author}
  {\bibfnamefont {P.~G.}\ \bibnamefont {Savvidis}}, \ and\ \bibinfo {author}
  {\bibfnamefont {J.~J.}\ \bibnamefont {Baumberg}},\ }\href {\doibase
  10.1103/PhysRevLett.116.106403} {\bibfield  {journal} {\bibinfo  {journal}
  {Phys. Rev. Lett.}\ }\textbf {\bibinfo {volume} {116}},\ \bibinfo {pages}
  {106403} (\bibinfo {year} {2016})}\BibitemShut {NoStop}%
\bibitem [{\citenamefont {Kavokin}\ \emph {et~al.}(2007)\citenamefont
  {Kavokin}, \citenamefont {Baumberg}, \citenamefont {Malpuech},\ and\
  \citenamefont {Laussy}}]{kavokin_microcavities_2007}%
  \BibitemOpen
  \bibfield  {author} {\bibinfo {author} {\bibfnamefont {A.~V.}\ \bibnamefont
  {Kavokin}}, \bibinfo {author} {\bibfnamefont {J.}~\bibnamefont {Baumberg}},
  \bibinfo {author} {\bibfnamefont {G.}~\bibnamefont {Malpuech}}, \ and\
  \bibinfo {author} {\bibfnamefont {F.~P.}\ \bibnamefont {Laussy}},\
  }\href@noop {} {\emph {\bibinfo {title} {Microcavities}}}\ (\bibinfo
  {publisher} {Oxford Univ. Press},\ \bibinfo {address} {Oxford},\ \bibinfo
  {year} {2007})\BibitemShut {NoStop}%
\bibitem [{\citenamefont {Savvidis}\ \emph {et~al.}(2000)\citenamefont
  {Savvidis}, \citenamefont {Baumberg}, \citenamefont {Stevenson},
  \citenamefont {Skolnick}, \citenamefont {Whittaker},\ and\ \citenamefont
  {Roberts}}]{savvidis_angle-resonant_2000}%
  \BibitemOpen
  \bibfield  {author} {\bibinfo {author} {\bibfnamefont {P.~G.}\ \bibnamefont
  {Savvidis}}, \bibinfo {author} {\bibfnamefont {J.~J.}\ \bibnamefont
  {Baumberg}}, \bibinfo {author} {\bibfnamefont {R.~M.}\ \bibnamefont
  {Stevenson}}, \bibinfo {author} {\bibfnamefont {M.~S.}\ \bibnamefont
  {Skolnick}}, \bibinfo {author} {\bibfnamefont {D.~M.}\ \bibnamefont
  {Whittaker}}, \ and\ \bibinfo {author} {\bibfnamefont {J.~S.}\ \bibnamefont
  {Roberts}},\ }\href {\doibase 10.1103/PhysRevLett.84.1547} {\bibfield
  {journal} {\bibinfo  {journal} {Phys. Rev. Lett.}\ }\textbf {\bibinfo
  {volume} {84}},\ \bibinfo {pages} {1547} (\bibinfo {year}
  {2000})}\BibitemShut {NoStop}%
\bibitem [{\citenamefont {Leyder}\ \emph {et~al.}(2007)\citenamefont {Leyder},
  \citenamefont {Romanelli}, \citenamefont {Karr}, \citenamefont {Giacobino},
  \citenamefont {Liew}, \citenamefont {Glazov}, \citenamefont {Kavokin},
  \citenamefont {Malpuech},\ and\ \citenamefont
  {Bramati}}]{leyder_observation_2007}%
  \BibitemOpen
  \bibfield  {author} {\bibinfo {author} {\bibfnamefont {C.}~\bibnamefont
  {Leyder}}, \bibinfo {author} {\bibfnamefont {M.}~\bibnamefont {Romanelli}},
  \bibinfo {author} {\bibfnamefont {J.~P.}\ \bibnamefont {Karr}}, \bibinfo
  {author} {\bibfnamefont {E.}~\bibnamefont {Giacobino}}, \bibinfo {author}
  {\bibfnamefont {T.~C.~H.}\ \bibnamefont {Liew}}, \bibinfo {author}
  {\bibfnamefont {M.~M.}\ \bibnamefont {Glazov}}, \bibinfo {author}
  {\bibfnamefont {A.~V.}\ \bibnamefont {Kavokin}}, \bibinfo {author}
  {\bibfnamefont {G.}~\bibnamefont {Malpuech}}, \ and\ \bibinfo {author}
  {\bibfnamefont {A.}~\bibnamefont {Bramati}},\ }\href {\doibase
  10.1038/nphys676} {\bibfield  {journal} {\bibinfo  {journal} {Nature Phys.}\
  }\textbf {\bibinfo {volume} {3}},\ \bibinfo {pages} {628} (\bibinfo {year}
  {2007})}\BibitemShut {NoStop}%
\bibitem [{\citenamefont {Lagoudakis}\ \emph {et~al.}(2009)\citenamefont
  {Lagoudakis}, \citenamefont {Ostatnický}, \citenamefont {Kavokin},
  \citenamefont {Rubo}, \citenamefont {André},\ and\ \citenamefont
  {Deveaud-Plédran}}]{lagoudakis_observation_2009}%
  \BibitemOpen
  \bibfield  {author} {\bibinfo {author} {\bibfnamefont {K.~G.}\ \bibnamefont
  {Lagoudakis}}, \bibinfo {author} {\bibfnamefont {T.}~\bibnamefont
  {Ostatnický}}, \bibinfo {author} {\bibfnamefont {A.~V.}\ \bibnamefont
  {Kavokin}}, \bibinfo {author} {\bibfnamefont {Y.~G.}\ \bibnamefont {Rubo}},
  \bibinfo {author} {\bibfnamefont {R.}~\bibnamefont {André}}, \ and\ \bibinfo
  {author} {\bibfnamefont {B.}~\bibnamefont {Deveaud-Plédran}},\ }\href
  {\doibase 10.1126/science.1177980} {\bibfield  {journal} {\bibinfo  {journal}
  {Science}\ }\textbf {\bibinfo {volume} {326}},\ \bibinfo {pages} {974 }
  (\bibinfo {year} {2009})}\BibitemShut {NoStop}%
\bibitem [{\citenamefont {Paraïso}\ \emph {et~al.}(2010)\citenamefont
  {Paraïso}, \citenamefont {Wouters}, \citenamefont {Léger}, \citenamefont
  {Morier-Genoud},\ and\ \citenamefont
  {Deveaud-Plédran}}]{paraiso_multistability_2010}%
  \BibitemOpen
  \bibfield  {author} {\bibinfo {author} {\bibfnamefont {T.~K.}\ \bibnamefont
  {Paraïso}}, \bibinfo {author} {\bibfnamefont {M.}~\bibnamefont {Wouters}},
  \bibinfo {author} {\bibfnamefont {Y.}~\bibnamefont {Léger}}, \bibinfo
  {author} {\bibfnamefont {F.}~\bibnamefont {Morier-Genoud}}, \ and\ \bibinfo
  {author} {\bibfnamefont {B.}~\bibnamefont {Deveaud-Plédran}},\ }\href
  {\doibase 10.1038/nmat2787} {\bibfield  {journal} {\bibinfo  {journal} {Nat.
  Mater.}\ }\textbf {\bibinfo {volume} {9}},\ \bibinfo {pages} {655} (\bibinfo
  {year} {2010})}\BibitemShut {NoStop}%
\bibitem [{\citenamefont {Abbarchi}\ \emph {et~al.}(2013)\citenamefont
  {Abbarchi}, \citenamefont {Amo}, \citenamefont {Sala}, \citenamefont
  {Solnyshkov}, \citenamefont {Flayac}, \citenamefont {Ferrier}, \citenamefont
  {Sagnes}, \citenamefont {Galopin}, \citenamefont {Lemaître}, \citenamefont
  {Malpuech},\ and\ \citenamefont {Bloch}}]{abbarchi_macroscopic_2013}%
  \BibitemOpen
  \bibfield  {author} {\bibinfo {author} {\bibfnamefont {M.}~\bibnamefont
  {Abbarchi}}, \bibinfo {author} {\bibfnamefont {A.}~\bibnamefont {Amo}},
  \bibinfo {author} {\bibfnamefont {V.~G.}\ \bibnamefont {Sala}}, \bibinfo
  {author} {\bibfnamefont {D.~D.}\ \bibnamefont {Solnyshkov}}, \bibinfo
  {author} {\bibfnamefont {H.}~\bibnamefont {Flayac}}, \bibinfo {author}
  {\bibfnamefont {L.}~\bibnamefont {Ferrier}}, \bibinfo {author} {\bibfnamefont
  {I.}~\bibnamefont {Sagnes}}, \bibinfo {author} {\bibfnamefont
  {E.}~\bibnamefont {Galopin}}, \bibinfo {author} {\bibfnamefont
  {A.}~\bibnamefont {Lemaître}}, \bibinfo {author} {\bibfnamefont
  {G.}~\bibnamefont {Malpuech}}, \ and\ \bibinfo {author} {\bibfnamefont
  {J.}~\bibnamefont {Bloch}},\ }\href {\doibase 10.1038/nphys2609} {\bibfield
  {journal} {\bibinfo  {journal} {Nat. Phys.}\ }\textbf {\bibinfo {volume}
  {9}},\ \bibinfo {pages} {275} (\bibinfo {year} {2013})}\BibitemShut {NoStop}%
\bibitem [{\citenamefont {Cristofolini}\ \emph {et~al.}(2013)\citenamefont
  {Cristofolini}, \citenamefont {Dreismann}, \citenamefont {Christmann},
  \citenamefont {Franchetti}, \citenamefont {Berloff}, \citenamefont {Tsotsis},
  \citenamefont {Hatzopoulos}, \citenamefont {Savvidis},\ and\ \citenamefont
  {Baumberg}}]{cristofolini_optical_2013}%
  \BibitemOpen
  \bibfield  {author} {\bibinfo {author} {\bibfnamefont {P.}~\bibnamefont
  {Cristofolini}}, \bibinfo {author} {\bibfnamefont {A.}~\bibnamefont
  {Dreismann}}, \bibinfo {author} {\bibfnamefont {G.}~\bibnamefont
  {Christmann}}, \bibinfo {author} {\bibfnamefont {G.}~\bibnamefont
  {Franchetti}}, \bibinfo {author} {\bibfnamefont {N.~G.}\ \bibnamefont
  {Berloff}}, \bibinfo {author} {\bibfnamefont {P.}~\bibnamefont {Tsotsis}},
  \bibinfo {author} {\bibfnamefont {Z.}~\bibnamefont {Hatzopoulos}}, \bibinfo
  {author} {\bibfnamefont {P.~G.}\ \bibnamefont {Savvidis}}, \ and\ \bibinfo
  {author} {\bibfnamefont {J.~J.}\ \bibnamefont {Baumberg}},\ }\href {\doibase
  10.1103/PhysRevLett.110.186403} {\bibfield  {journal} {\bibinfo  {journal}
  {Phys. Rev. Lett.}\ }\textbf {\bibinfo {volume} {110}},\ \bibinfo {pages}
  {186403} (\bibinfo {year} {2013})}\BibitemShut {NoStop}%
\bibitem [{\citenamefont {Sala}\ \emph {et~al.}(2015)\citenamefont {Sala},
  \citenamefont {Solnyshkov}, \citenamefont {Carusotto}, \citenamefont
  {Jacqmin}, \citenamefont {Lemaître}, \citenamefont {Terças}, \citenamefont
  {Nalitov}, \citenamefont {Abbarchi}, \citenamefont {Galopin}, \citenamefont
  {Sagnes} \emph {et~al.}}]{sala_spin-orbit_2015}%
  \BibitemOpen
  \bibfield  {author} {\bibinfo {author} {\bibfnamefont {V.~G.}\ \bibnamefont
  {Sala}}, \bibinfo {author} {\bibfnamefont {D.~D.}\ \bibnamefont
  {Solnyshkov}}, \bibinfo {author} {\bibfnamefont {I.}~\bibnamefont
  {Carusotto}}, \bibinfo {author} {\bibfnamefont {T.}~\bibnamefont {Jacqmin}},
  \bibinfo {author} {\bibfnamefont {A.}~\bibnamefont {Lemaître}}, \bibinfo
  {author} {\bibfnamefont {H.}~\bibnamefont {Terças}}, \bibinfo {author}
  {\bibfnamefont {A.}~\bibnamefont {Nalitov}}, \bibinfo {author} {\bibfnamefont
  {M.}~\bibnamefont {Abbarchi}}, \bibinfo {author} {\bibfnamefont
  {E.}~\bibnamefont {Galopin}}, \bibinfo {author} {\bibfnamefont
  {I.}~\bibnamefont {Sagnes}},  \emph {et~al.},\ }\href {\doibase
  10.1103/PhysRevX.5.011034} {\bibfield  {journal} {\bibinfo  {journal} {Phys.
  Rev. X}\ }\textbf {\bibinfo {volume} {5}},\ \bibinfo {pages} {011034}
  (\bibinfo {year} {2015})}\BibitemShut {NoStop}%
\bibitem [{\citenamefont {Dufferwiel}\ \emph {et~al.}(2015)\citenamefont
  {Dufferwiel}, \citenamefont {Li}, \citenamefont {Cancellieri}, \citenamefont
  {Giriunas}, \citenamefont {Trichet}, \citenamefont {Whittaker}, \citenamefont
  {Walker}, \citenamefont {Fras}, \citenamefont {Clarke}, \citenamefont
  {Smith}, \citenamefont {Skolnick},\ and\ \citenamefont
  {Krizhanovskii}}]{dufferwiel_spin_2015}%
  \BibitemOpen
  \bibfield  {author} {\bibinfo {author} {\bibfnamefont {S.}~\bibnamefont
  {Dufferwiel}}, \bibinfo {author} {\bibfnamefont {F.}~\bibnamefont {Li}},
  \bibinfo {author} {\bibfnamefont {E.}~\bibnamefont {Cancellieri}}, \bibinfo
  {author} {\bibfnamefont {L.}~\bibnamefont {Giriunas}}, \bibinfo {author}
  {\bibfnamefont {A.~A.~P.}\ \bibnamefont {Trichet}}, \bibinfo {author}
  {\bibfnamefont {D.~M.}\ \bibnamefont {Whittaker}}, \bibinfo {author}
  {\bibfnamefont {P.~M.}\ \bibnamefont {Walker}}, \bibinfo {author}
  {\bibfnamefont {F.}~\bibnamefont {Fras}}, \bibinfo {author} {\bibfnamefont
  {E.}~\bibnamefont {Clarke}}, \bibinfo {author} {\bibfnamefont {J.~M.}\
  \bibnamefont {Smith}}, \bibinfo {author} {\bibfnamefont {M.~S.}\ \bibnamefont
  {Skolnick}}, \ and\ \bibinfo {author} {\bibfnamefont {D.~N.}\ \bibnamefont
  {Krizhanovskii}},\ }\href {\doibase 10.1103/PhysRevLett.115.246401}
  {\bibfield  {journal} {\bibinfo  {journal} {Phys. Rev. Lett.}\ }\textbf
  {\bibinfo {volume} {115}},\ \bibinfo {pages} {246401} (\bibinfo {year}
  {2015})}\BibitemShut {NoStop}%
\bibitem [{\citenamefont {Amo}\ \emph {et~al.}(2010)\citenamefont {Amo},
  \citenamefont {Liew}, \citenamefont {Adrados}, \citenamefont {Houdre},
  \citenamefont {Giacobino}, \citenamefont {Kavokin},\ and\ \citenamefont
  {Bramati}}]{amo_exciton-polariton_2010}%
  \BibitemOpen
  \bibfield  {author} {\bibinfo {author} {\bibfnamefont {A.}~\bibnamefont
  {Amo}}, \bibinfo {author} {\bibfnamefont {T.~C.~H.}\ \bibnamefont {Liew}},
  \bibinfo {author} {\bibfnamefont {C.}~\bibnamefont {Adrados}}, \bibinfo
  {author} {\bibfnamefont {R.}~\bibnamefont {Houdre}}, \bibinfo {author}
  {\bibfnamefont {E.}~\bibnamefont {Giacobino}}, \bibinfo {author}
  {\bibfnamefont {A.~V.}\ \bibnamefont {Kavokin}}, \ and\ \bibinfo {author}
  {\bibfnamefont {A.}~\bibnamefont {Bramati}},\ }\href {\doibase
  10.1038/nphoton.2010.79} {\bibfield  {journal} {\bibinfo  {journal} {Nature
  Photon.}\ }\textbf {\bibinfo {volume} {4}},\ \bibinfo {pages} {361} (\bibinfo
  {year} {2010})}\BibitemShut {NoStop}%
\bibitem [{\citenamefont {Ballarini}\ \emph {et~al.}(2013)\citenamefont
  {Ballarini}, \citenamefont {De~Giorgi}, \citenamefont {Cancellieri},
  \citenamefont {Houdré}, \citenamefont {Giacobino}, \citenamefont
  {Cingolani}, \citenamefont {Bramati}, \citenamefont {Gigli},\ and\
  \citenamefont {Sanvitto}}]{ballarini_all-optical_2013}%
  \BibitemOpen
  \bibfield  {author} {\bibinfo {author} {\bibfnamefont {D.}~\bibnamefont
  {Ballarini}}, \bibinfo {author} {\bibfnamefont {M.}~\bibnamefont
  {De~Giorgi}}, \bibinfo {author} {\bibfnamefont {E.}~\bibnamefont
  {Cancellieri}}, \bibinfo {author} {\bibfnamefont {R.}~\bibnamefont
  {Houdré}}, \bibinfo {author} {\bibfnamefont {E.}~\bibnamefont {Giacobino}},
  \bibinfo {author} {\bibfnamefont {R.}~\bibnamefont {Cingolani}}, \bibinfo
  {author} {\bibfnamefont {A.}~\bibnamefont {Bramati}}, \bibinfo {author}
  {\bibfnamefont {G.}~\bibnamefont {Gigli}}, \ and\ \bibinfo {author}
  {\bibfnamefont {D.}~\bibnamefont {Sanvitto}},\ }\href {\doibase
  10.1038/ncomms2734} {\bibfield  {journal} {\bibinfo  {journal} {Nat.
  Commun.}\ }\textbf {\bibinfo {volume} {4}},\ \bibinfo {pages} {1778}
  (\bibinfo {year} {2013})}\BibitemShut {NoStop}%
\bibitem [{\citenamefont {Cerna}\ \emph {et~al.}(2013)\citenamefont {Cerna},
  \citenamefont {Léger}, \citenamefont {Paraïso}, \citenamefont {Wouters},
  \citenamefont {Morier-Genoud}, \citenamefont {Portella-Oberli},\ and\
  \citenamefont {Deveaud}}]{cerna_ultrafast_2013}%
  \BibitemOpen
  \bibfield  {author} {\bibinfo {author} {\bibfnamefont {R.}~\bibnamefont
  {Cerna}}, \bibinfo {author} {\bibfnamefont {Y.}~\bibnamefont {Léger}},
  \bibinfo {author} {\bibfnamefont {T.~K.}\ \bibnamefont {Paraïso}}, \bibinfo
  {author} {\bibfnamefont {M.}~\bibnamefont {Wouters}}, \bibinfo {author}
  {\bibfnamefont {F.}~\bibnamefont {Morier-Genoud}}, \bibinfo {author}
  {\bibfnamefont {M.~T.}\ \bibnamefont {Portella-Oberli}}, \ and\ \bibinfo
  {author} {\bibfnamefont {B.}~\bibnamefont {Deveaud}},\ }\href {\doibase
  10.1038/ncomms3008} {\bibfield  {journal} {\bibinfo  {journal} {Nat.
  Commun.}\ }\textbf {\bibinfo {volume} {4}},\ \bibinfo {pages} {2008}
  (\bibinfo {year} {2013})}\BibitemShut {NoStop}%
\bibitem [{\citenamefont {Nguyen}\ \emph {et~al.}(2013)\citenamefont {Nguyen},
  \citenamefont {Vishnevsky}, \citenamefont {Sturm}, \citenamefont {Tanese},
  \citenamefont {Solnyshkov}, \citenamefont {Galopin}, \citenamefont
  {Lemaître}, \citenamefont {Sagnes}, \citenamefont {Amo}, \citenamefont
  {Malpuech},\ and\ \citenamefont {Bloch}}]{nguyen_realization_2013}%
  \BibitemOpen
  \bibfield  {author} {\bibinfo {author} {\bibfnamefont {H.~S.}\ \bibnamefont
  {Nguyen}}, \bibinfo {author} {\bibfnamefont {D.}~\bibnamefont {Vishnevsky}},
  \bibinfo {author} {\bibfnamefont {C.}~\bibnamefont {Sturm}}, \bibinfo
  {author} {\bibfnamefont {D.}~\bibnamefont {Tanese}}, \bibinfo {author}
  {\bibfnamefont {D.}~\bibnamefont {Solnyshkov}}, \bibinfo {author}
  {\bibfnamefont {E.}~\bibnamefont {Galopin}}, \bibinfo {author} {\bibfnamefont
  {A.}~\bibnamefont {Lemaître}}, \bibinfo {author} {\bibfnamefont
  {I.}~\bibnamefont {Sagnes}}, \bibinfo {author} {\bibfnamefont
  {A.}~\bibnamefont {Amo}}, \bibinfo {author} {\bibfnamefont {G.}~\bibnamefont
  {Malpuech}}, \ and\ \bibinfo {author} {\bibfnamefont {J.}~\bibnamefont
  {Bloch}},\ }\href {\doibase 10.1103/PhysRevLett.110.236601} {\bibfield
  {journal} {\bibinfo  {journal} {Phys. Rev. Lett.}\ }\textbf {\bibinfo
  {volume} {110}},\ \bibinfo {pages} {236601} (\bibinfo {year}
  {2013})}\BibitemShut {NoStop}%
\bibitem [{\citenamefont {Dreismann}\ \emph {et~al.}(2016)\citenamefont
  {Dreismann}, \citenamefont {Ohadi}, \citenamefont {del Valle-Inclan~Redondo},
  \citenamefont {Balili}, \citenamefont {Rubo}, \citenamefont {Tsintzos},
  \citenamefont {Deligeorgis}, \citenamefont {Hatzopoulos}, \citenamefont
  {Savvidis},\ and\ \citenamefont {Baumberg}}]{dreismann_sub-femtojoule_2016}%
  \BibitemOpen
  \bibfield  {author} {\bibinfo {author} {\bibfnamefont {A.}~\bibnamefont
  {Dreismann}}, \bibinfo {author} {\bibfnamefont {H.}~\bibnamefont {Ohadi}},
  \bibinfo {author} {\bibfnamefont {Y.}~\bibnamefont {del
  Valle-Inclan~Redondo}}, \bibinfo {author} {\bibfnamefont {R.}~\bibnamefont
  {Balili}}, \bibinfo {author} {\bibfnamefont {Y.~G.}\ \bibnamefont {Rubo}},
  \bibinfo {author} {\bibfnamefont {S.~I.}\ \bibnamefont {Tsintzos}}, \bibinfo
  {author} {\bibfnamefont {G.}~\bibnamefont {Deligeorgis}}, \bibinfo {author}
  {\bibfnamefont {Z.}~\bibnamefont {Hatzopoulos}}, \bibinfo {author}
  {\bibfnamefont {P.~G.}\ \bibnamefont {Savvidis}}, \ and\ \bibinfo {author}
  {\bibfnamefont {J.~J.}\ \bibnamefont {Baumberg}},\ }\href {\doibase
  10.1038/nmat4722} {\bibfield  {journal} {\bibinfo  {journal} {Nat. Mater.}\
  }\textbf {\bibinfo {volume} {15}},\ \bibinfo {pages} {1074} (\bibinfo {year}
  {2016})}\BibitemShut {NoStop}%
\bibitem [{\citenamefont {Wertz}\ \emph {et~al.}(2010)\citenamefont {Wertz},
  \citenamefont {Ferrier}, \citenamefont {Solnyshkov}, \citenamefont {Johne},
  \citenamefont {Sanvitto}, \citenamefont {Lemaître}, \citenamefont {Sagnes},
  \citenamefont {Grousson}, \citenamefont {Kavokin}, \citenamefont {Senellart},
  \citenamefont {Malpuech},\ and\ \citenamefont
  {Bloch}}]{wertz_spontaneous_2010}%
  \BibitemOpen
  \bibfield  {author} {\bibinfo {author} {\bibfnamefont {E.}~\bibnamefont
  {Wertz}}, \bibinfo {author} {\bibfnamefont {L.}~\bibnamefont {Ferrier}},
  \bibinfo {author} {\bibfnamefont {D.~D.}\ \bibnamefont {Solnyshkov}},
  \bibinfo {author} {\bibfnamefont {R.}~\bibnamefont {Johne}}, \bibinfo
  {author} {\bibfnamefont {D.}~\bibnamefont {Sanvitto}}, \bibinfo {author}
  {\bibfnamefont {A.}~\bibnamefont {Lemaître}}, \bibinfo {author}
  {\bibfnamefont {I.}~\bibnamefont {Sagnes}}, \bibinfo {author} {\bibfnamefont
  {R.}~\bibnamefont {Grousson}}, \bibinfo {author} {\bibfnamefont {A.~V.}\
  \bibnamefont {Kavokin}}, \bibinfo {author} {\bibfnamefont {P.}~\bibnamefont
  {Senellart}}, \bibinfo {author} {\bibfnamefont {G.}~\bibnamefont {Malpuech}},
  \ and\ \bibinfo {author} {\bibfnamefont {J.}~\bibnamefont {Bloch}},\ }\href
  {\doibase 10.1038/nphys1750} {\bibfield  {journal} {\bibinfo  {journal}
  {Nature Phys.}\ }\textbf {\bibinfo {volume} {6}},\ \bibinfo {pages} {860}
  (\bibinfo {year} {2010})}\BibitemShut {NoStop}%
\bibitem [{\citenamefont {Tosi}\ \emph {et~al.}(2012)\citenamefont {Tosi},
  \citenamefont {Christmann}, \citenamefont {Berloff}, \citenamefont {Tsotsis},
  \citenamefont {Gao}, \citenamefont {Hatzopoulos}, \citenamefont {Savvidis},\
  and\ \citenamefont {Baumberg}}]{tosi_sculpting_2012}%
  \BibitemOpen
  \bibfield  {author} {\bibinfo {author} {\bibfnamefont {G.}~\bibnamefont
  {Tosi}}, \bibinfo {author} {\bibfnamefont {G.}~\bibnamefont {Christmann}},
  \bibinfo {author} {\bibfnamefont {N.~G.}\ \bibnamefont {Berloff}}, \bibinfo
  {author} {\bibfnamefont {P.}~\bibnamefont {Tsotsis}}, \bibinfo {author}
  {\bibfnamefont {T.}~\bibnamefont {Gao}}, \bibinfo {author} {\bibfnamefont
  {Z.}~\bibnamefont {Hatzopoulos}}, \bibinfo {author} {\bibfnamefont {P.~G.}\
  \bibnamefont {Savvidis}}, \ and\ \bibinfo {author} {\bibfnamefont {J.~J.}\
  \bibnamefont {Baumberg}},\ }\href {\doibase 10.1038/nphys2182} {\bibfield
  {journal} {\bibinfo  {journal} {Nature Phys.}\ }\textbf {\bibinfo {volume}
  {8}},\ \bibinfo {pages} {190} (\bibinfo {year} {2012})}\BibitemShut {NoStop}%
\bibitem [{\citenamefont {Askitopoulos}\ \emph {et~al.}(2013)\citenamefont
  {Askitopoulos}, \citenamefont {Ohadi}, \citenamefont {Kavokin}, \citenamefont
  {Hatzopoulos}, \citenamefont {Savvidis},\ and\ \citenamefont
  {Lagoudakis}}]{askitopoulos_polariton_2013}%
  \BibitemOpen
  \bibfield  {author} {\bibinfo {author} {\bibfnamefont {A.}~\bibnamefont
  {Askitopoulos}}, \bibinfo {author} {\bibfnamefont {H.}~\bibnamefont {Ohadi}},
  \bibinfo {author} {\bibfnamefont {A.~V.}\ \bibnamefont {Kavokin}}, \bibinfo
  {author} {\bibfnamefont {Z.}~\bibnamefont {Hatzopoulos}}, \bibinfo {author}
  {\bibfnamefont {P.~G.}\ \bibnamefont {Savvidis}}, \ and\ \bibinfo {author}
  {\bibfnamefont {P.~G.}\ \bibnamefont {Lagoudakis}},\ }\href {\doibase
  10.1103/PhysRevB.88.041308} {\bibfield  {journal} {\bibinfo  {journal} {Phys.
  Rev. B}\ }\textbf {\bibinfo {volume} {88}},\ \bibinfo {pages} {041308}
  (\bibinfo {year} {2013})}\BibitemShut {NoStop}%
\bibitem [{\citenamefont {Aleiner}\ \emph {et~al.}(2012)\citenamefont
  {Aleiner}, \citenamefont {Altshuler},\ and\ \citenamefont
  {Rubo}}]{aleiner_radiative_2012}%
  \BibitemOpen
  \bibfield  {author} {\bibinfo {author} {\bibfnamefont {I.~L.}\ \bibnamefont
  {Aleiner}}, \bibinfo {author} {\bibfnamefont {B.~L.}\ \bibnamefont
  {Altshuler}}, \ and\ \bibinfo {author} {\bibfnamefont {Y.~G.}\ \bibnamefont
  {Rubo}},\ }\href {\doibase 10.1103/PhysRevB.85.121301} {\bibfield  {journal}
  {\bibinfo  {journal} {Phys. Rev. B}\ }\textbf {\bibinfo {volume} {85}},\
  \bibinfo {pages} {121301} (\bibinfo {year} {2012})}\BibitemShut {NoStop}%
\bibitem [{\citenamefont {Keeling}\ and\ \citenamefont
  {Berloff}(2008)}]{keeling_spontaneous_2008}%
  \BibitemOpen
  \bibfield  {author} {\bibinfo {author} {\bibfnamefont {J.}~\bibnamefont
  {Keeling}}\ and\ \bibinfo {author} {\bibfnamefont {N.~G.}\ \bibnamefont
  {Berloff}},\ }\href {\doibase 10.1103/PhysRevLett.100.250401} {\bibfield
  {journal} {\bibinfo  {journal} {Phys. Rev. Lett.}\ }\textbf {\bibinfo
  {volume} {100}},\ \bibinfo {pages} {250401} (\bibinfo {year}
  {2008})}\BibitemShut {NoStop}%
\bibitem [{\citenamefont {Lagoudakis}\ \emph {et~al.}(2010)\citenamefont
  {Lagoudakis}, \citenamefont {Pietka}, \citenamefont {Wouters}, \citenamefont
  {André},\ and\ \citenamefont {Deveaud-Plédran}}]{lagoudakis_coherent_2010}%
  \BibitemOpen
  \bibfield  {author} {\bibinfo {author} {\bibfnamefont {K.~G.}\ \bibnamefont
  {Lagoudakis}}, \bibinfo {author} {\bibfnamefont {B.}~\bibnamefont {Pietka}},
  \bibinfo {author} {\bibfnamefont {M.}~\bibnamefont {Wouters}}, \bibinfo
  {author} {\bibfnamefont {R.}~\bibnamefont {André}}, \ and\ \bibinfo {author}
  {\bibfnamefont {B.}~\bibnamefont {Deveaud-Plédran}},\ }\href {\doibase
  10.1103/PhysRevLett.105.120403} {\bibfield  {journal} {\bibinfo  {journal}
  {Phys. Rev. Lett.}\ }\textbf {\bibinfo {volume} {105}},\ \bibinfo {pages}
  {120403} (\bibinfo {year} {2010})}\BibitemShut {NoStop}%
\bibitem [{\citenamefont {Wouters}(2008)}]{wouters_synchronized_2008}%
  \BibitemOpen
  \bibfield  {author} {\bibinfo {author} {\bibfnamefont {M.}~\bibnamefont
  {Wouters}},\ }\href {\doibase 10.1103/PhysRevB.77.121302} {\bibfield
  {journal} {\bibinfo  {journal} {Physical Review B}\ }\textbf {\bibinfo
  {volume} {77}},\ \bibinfo {pages} {121302} (\bibinfo {year}
  {2008})}\BibitemShut {NoStop}%
\bibitem [{\citenamefont {Borgh}\ \emph {et~al.}(2010)\citenamefont {Borgh},
  \citenamefont {Keeling},\ and\ \citenamefont {Berloff}}]{borgh_spatial_2010}%
  \BibitemOpen
  \bibfield  {author} {\bibinfo {author} {\bibfnamefont {M.~O.}\ \bibnamefont
  {Borgh}}, \bibinfo {author} {\bibfnamefont {J.}~\bibnamefont {Keeling}}, \
  and\ \bibinfo {author} {\bibfnamefont {N.~G.}\ \bibnamefont {Berloff}},\
  }\href {\doibase 10.1103/PhysRevB.81.235302} {\bibfield  {journal} {\bibinfo
  {journal} {Physical Review B}\ }\textbf {\bibinfo {volume} {81}},\ \bibinfo
  {pages} {235302} (\bibinfo {year} {2010})}\BibitemShut {NoStop}%
\bibitem [{\citenamefont {Matousek}(2010)}]{matousek_thirty-three_2010}%
  \BibitemOpen
  \bibfield  {author} {\bibinfo {author} {\bibfnamefont {J.}~\bibnamefont
  {Matousek}},\ }\href@noop {} {\emph {\bibinfo {title} {Thirty-three
  {Miniatures}: {Mathematical} and {Algorithmic} {Applications} of {Linear}
  {Algebra}}}}\ (\bibinfo  {publisher} {American Mathematical Society},\
  \bibinfo {address} {Providence, R.I},\ \bibinfo {year} {2010})\
  Chap.~\bibinfo {chapter} {22}\BibitemShut {NoStop}%
\bibitem [{\citenamefont {Tsotsis}\ \emph {et~al.}(2012)\citenamefont
  {Tsotsis}, \citenamefont {Eldridge}, \citenamefont {Gao}, \citenamefont
  {Tsintzos}, \citenamefont {Hatzopoulos},\ and\ \citenamefont
  {Savvidis}}]{tsotsis_lasing_2012}%
  \BibitemOpen
  \bibfield  {author} {\bibinfo {author} {\bibfnamefont {P.}~\bibnamefont
  {Tsotsis}}, \bibinfo {author} {\bibfnamefont {P.~S.}\ \bibnamefont
  {Eldridge}}, \bibinfo {author} {\bibfnamefont {T.}~\bibnamefont {Gao}},
  \bibinfo {author} {\bibfnamefont {S.~I.}\ \bibnamefont {Tsintzos}}, \bibinfo
  {author} {\bibfnamefont {Z.}~\bibnamefont {Hatzopoulos}}, \ and\ \bibinfo
  {author} {\bibfnamefont {P.~G.}\ \bibnamefont {Savvidis}},\ }\href {\doibase
  10.1088/1367-2630/14/2/023060} {\bibfield  {journal} {\bibinfo  {journal} {N.
  J. Phys.}\ }\textbf {\bibinfo {volume} {14}},\ \bibinfo {pages} {023060}
  (\bibinfo {year} {2012})}\BibitemShut {NoStop}%
\bibitem [{\citenamefont {Wouters}\ and\ \citenamefont
  {Carusotto}(2007)}]{wouters_excitations_2007}%
  \BibitemOpen
  \bibfield  {author} {\bibinfo {author} {\bibfnamefont {M.}~\bibnamefont
  {Wouters}}\ and\ \bibinfo {author} {\bibfnamefont {I.}~\bibnamefont
  {Carusotto}},\ }\href {\doibase 10.1103/PhysRevLett.99.140402} {\bibfield
  {journal} {\bibinfo  {journal} {Phys. Rev. Lett.}\ }\textbf {\bibinfo
  {volume} {99}},\ \bibinfo {pages} {140402} (\bibinfo {year}
  {2007})}\BibitemShut {NoStop}%
\bibitem [{\citenamefont {Wouters}\ \emph {et~al.}(2010)\citenamefont
  {Wouters}, \citenamefont {Liew},\ and\ \citenamefont
  {Savona}}]{wouters_energy_2010}%
  \BibitemOpen
  \bibfield  {author} {\bibinfo {author} {\bibfnamefont {M.}~\bibnamefont
  {Wouters}}, \bibinfo {author} {\bibfnamefont {T.~C.~H.}\ \bibnamefont
  {Liew}}, \ and\ \bibinfo {author} {\bibfnamefont {V.}~\bibnamefont
  {Savona}},\ }\href {\doibase 10.1103/PhysRevB.82.245315} {\bibfield
  {journal} {\bibinfo  {journal} {Phys. Rev. B}\ }\textbf {\bibinfo {volume}
  {82}},\ \bibinfo {pages} {245315} (\bibinfo {year} {2010})}\BibitemShut
  {NoStop}%
\end{thebibliography}%


\clearpage
\pagebreak
\onecolumngrid
\newgeometry{textwidth=390pt,textheight=592pt,footskip=1in}

\begin{center}
\textbf{\large Supplemental Information}
\end{center}
\setcounter{equation}{0}
\setcounter{figure}{0}
\setcounter{table}{0}
\setcounter{page}{1}
\makeatletter
\renewcommand{\theequation}{S\arabic{equation}}
\renewcommand{\thefigure}{S\arabic{figure}}
\renewcommand\thesection{\arabic{section}}
\graphicspath{{./SIfigs/}}

\section{Sample and experimental techniques}
\label{si:sample} The cavity top (bottom) distributed
Bragg reflector (DBR) is made of 32 (35) pairs of
Al$_{0.15}$Ga$_{0.85}$As/AlAs layers of \SI{57.2}{\nm}/\SI{65.4}{\nm}. Four
sets of three \SI{10}{nm} GaAs quantum wells (QW) separated by \SI{10}{nm}
thick layers of Al$_{0.3}$Ga$_{0.7}$As are placed at the maxima of the
cavity light field.  The 5$\lambda$/2 (\SI{583}{\nm}) cavity is made of
Al$_{0.3}$Ga$_{0.7}$As. The sample shows condensation under non-resonant
excitation~\cite{tsotsis_lasing_2012}.

The quasi continuous
wave pump is a single-mode Ti:Sapphire laser tuned to the first Bragg mode
\SI{\sim 100}{\milli\electronvolt} above the condensate energy, and is
linearly polarized. In order to drive condensate formation on short time
scales, the pump is amplitude-modulated by an acousto-optic modulator. A
spatial light modulator (SLM) is used to spatially pattern the pump beam
into a square lattice.
 A 0.4 NA objective is used for imaging the pattern onto the
sample. A cooled CCD and a \SI{0.55}{\meter} spectrometer is used for
imaging and energy resolving the emission. Polarization is analysed
using a quarter-waveplate and a Wollaston prism in front of the camera. The
CCD and the pump laser are electronically synchronized.
For each realization of the experiment, the sample is exposed to
microsecond-long pump pulses, and the final spin state is measured. Many
realizations are measured to build statistics.

\section{Momentum, real-space and potential landscape}
\label{si:potlands}

\begin{figure*}
	\centering
	\begin{minipage}[t]{.48\textwidth}
		\centering
		\includegraphics[width=1.\linewidth]{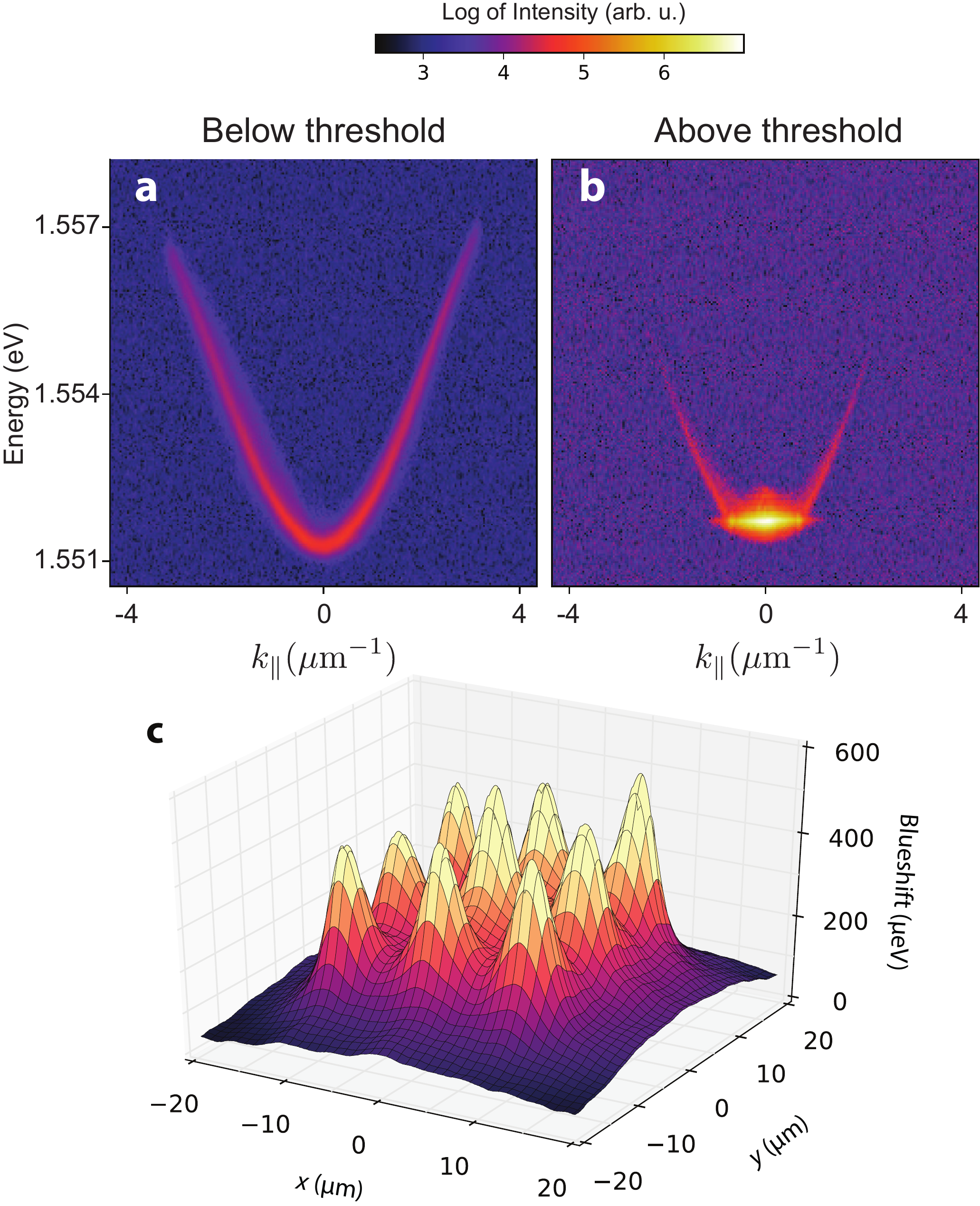}
		\caption{\captit{Momentum and potential landscape.} \capind{a} In-plane
			energy-resolved momentum below threshold
			in-plane momentum, and \capind{b} above threshold. \capind{c} The
			pump-induced potential.}\label{fig:potential}
	\end{minipage}\hfill
	\begin{minipage}[t]{.48\textwidth}
		\centering
		\includegraphics[width=1.\linewidth]{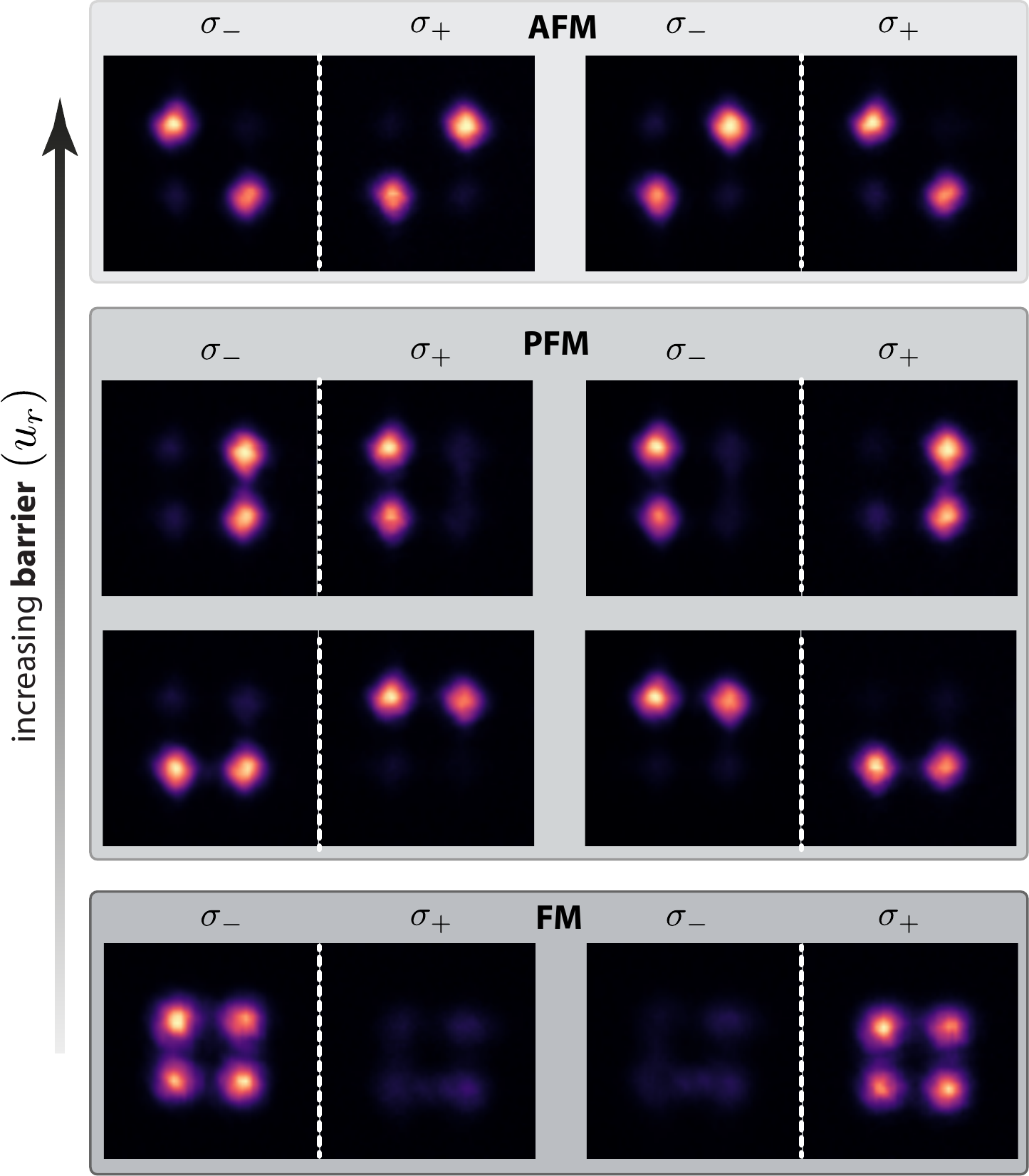}
		\caption{\captit{Real-space polarization-resolved intensity.} Spin-up
			($\sigma_+$) and
			spin-down ($\sigma_-$) intensities of steady states as barrier is increased.}\label{fig:realization}
	\end{minipage}
\end{figure*}

Fig.~\ref{fig:potential}a shows the logarithmic scale of the momentum space
intensity at below condensation threshold. The cavity detuning here is
\SIrange{-2}{-3}{\meV}. As the pumping power increases we observe a continuous
blueshift of up to \SI{\sim300}{\ueV} until the condensates reach the spin
bifurcation threshold. At this threshold, all 4 lattice sites condense to the
ground state of the dispersion at a single energy to within the resolution of
our spectrometer (Fig.~\ref{fig:potential}b). Fig.~\ref{fig:potential}c shows
the potential landscape. At the pump spots the blueshift is \SI{\sim600}{\ueV}
at $u_r=1$, whereas at the saddle points it drops to \SI{\sim300}{\ueV}.  The
confining potential at the saddle points is \SI{\sim80}{\ueV}.

Fig.~\ref{fig:realization} shows the $\sigma_+$ and $\sigma_-$ polarized
intensity of the condensate chain stable states at FM, PFM and AFM regimes.
The minimum circular polarization of the condensates is 75\%.

\section{Feedback search}
\label{si:feedback}

\begin{figure*}
	\centering
	\includegraphics[width=1\textwidth]{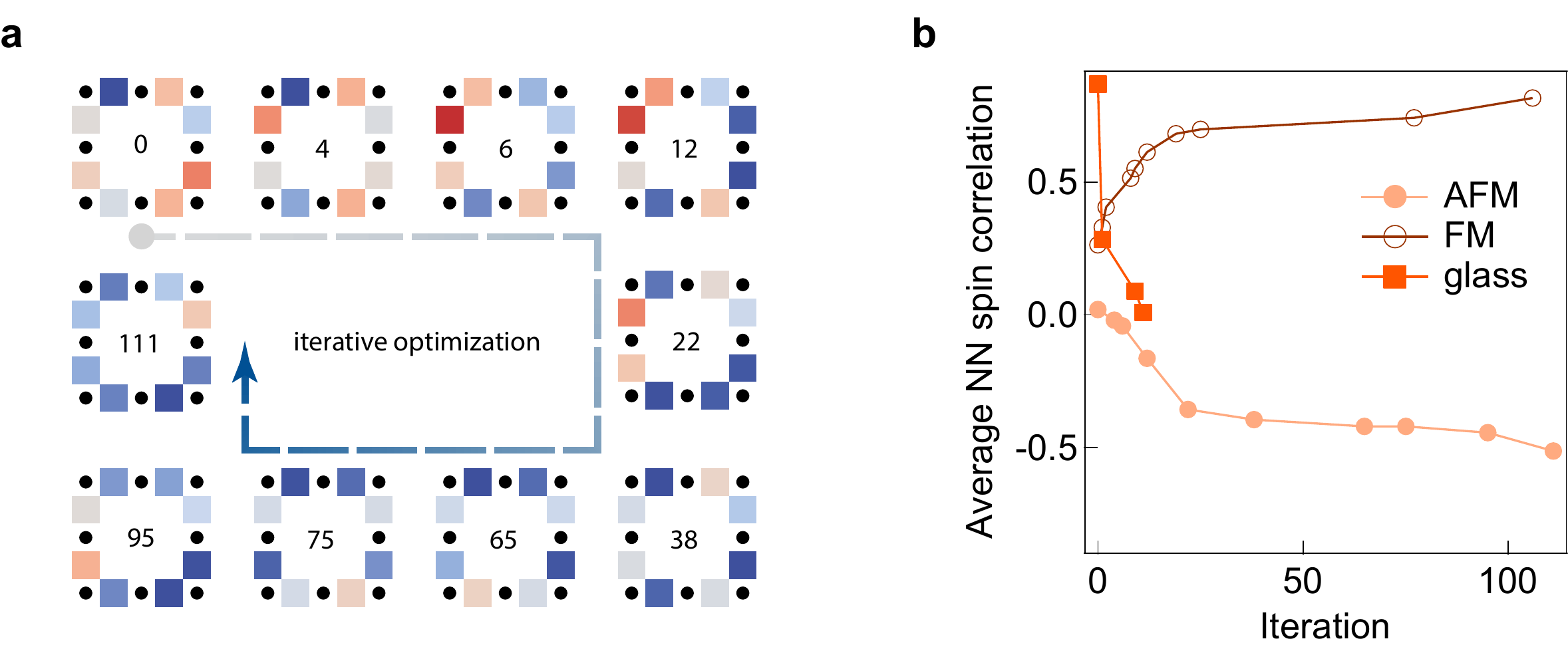}
	\caption{\captit{Feedback search} \capind{a} Example of iterative search for AFM spin chain, showing
	the spin correlation of the bonds (indices show iteration number). Blue
	links denote AFM bonds and reds are FM bonds and colour scales are as in
	Fig.~2. Indexes show the iteration number. \capind{b} Average
	nearest-neighbor spin correlation as a function of search iteration step
	for three search criteria of FM, glass and AFM target phases.}\label{fig:siFeedback}
\end{figure*}
The search algorithm starts with equal intensities across the array and randomly
changes each pump spot intensity by up to of 10\% on each step. For each
iteration we record 100 realizations and calculate the average NN spin
correlation $\bar{C}_{NN}$ as a figure of merit. To seek an AFM chain, we look
for configurations where $\bar{C}_{NN}$ is minimized, retaining intensity
patterns that result in a smaller $\bar{C}_{NN}$. A typical example of this
iterative process is shown in Fig.~\ref{fig:siFeedback}a. With \SI{1}{\second}
per iteration the search time for desired spin chains is only minutes. For FM
chains $\bar{C}_{NN}$ is instead maximized, while for the glass state
$\vert\bar{C}_{NN}\vert$ is minimized (Fig.~\ref{fig:siFeedback}b).

In principle, the correction mechanism is scalable to larger lattices, and only
needs to be applied once. The limiting factors are in fact the total pump
intensity and the field of view of the objective. The efficiency of our optical setup
is currently ~20\%, meaning that for 1W pump laser we get \SI{200}{\milli\watt} patterned power
on the sample. For condensation we require \SI{\sim 5}{\milli\watt} per spot (4 spots for a site),
which is currently limiting us to a 8$\times$8-condensate lattice (\SI{2}{\watt} pump laser). The other
limiting factor is the field of view of the objective. Currently each site is
\SI{\sim 10}{\um} and in order to get the required pump spot resolution of \SI{\sim 1}{\um}, we need
to work with high numerical aperture objectives, which limit our field of view to
\SI{\sim1}{\milli\meter} diameters, effectively limiting us to 100$\times$100-condensate lattices, but large
enough to study intriguing experiments.

	\section{Monte-Carlo simulations}\label{si:simulations}
	\subsection{zero-dimensional}

	We perform Monte-Carlo simulations while $J$ is varied. We perform 2000
	realizations that lasts \SI{4}{\nano\second} with random initial conditions
	for each $J$ and calculate the correlations of the final state $s_z$ of the
	condensate realizations. Fig.~\ref{fig:0D} shows the correlation of the
	sides and diagonal condensates versus the Josephson coupling strength $J$ at
	$P=1.1 P_c$. We observe a qualitatively similar behavior as the 2D
	simulations: as $J$ reduces we go from FM to PFM to AFM phases. The
	parameters used for 0D simulations were
	$W=\SI{0.2}{\ps^{-1}}$; $\Gamma=\SI{.1}{\ps^{-1}}$;
	$\epsilon=\SI{0.04}{\ps^{-1}}$; $\gamma=0.2\epsilon$;
	$\alpha_1=\SI{0.01}{\ps^{-1}}$; $\alpha_2=-0.5\alpha_1$;
	$\eta=\SI{0.02}{\ps^{-1}}$.

	\subsection{2-dimensional}

	\begin{figure*}
		\centering
		\begin{minipage}[t]{.8\textwidth}
			\centering
			\includegraphics[width=.8\textwidth]{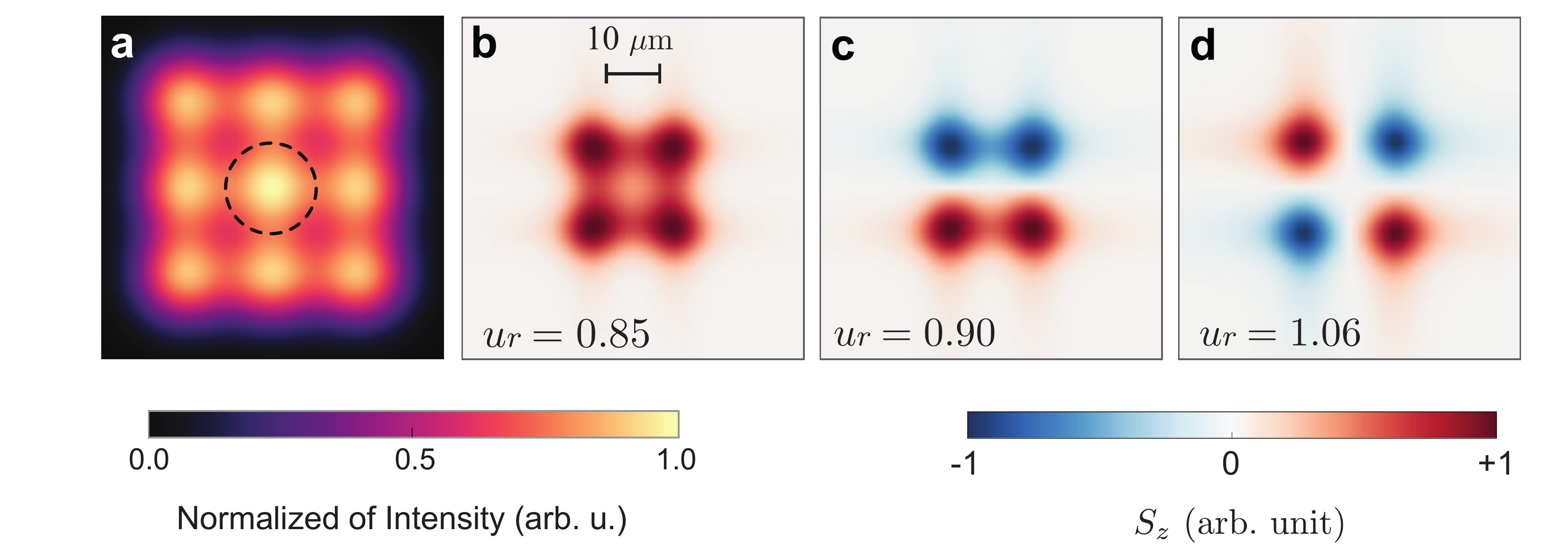} \caption{\captit{Steady states 2D
				simulations.} \capind{a} Pump intensity pattern at $u_r=1.08$. The
				variable pump spot intensity is marked by a dashed circle.  \capind{b-d}
				Condensate spin $S_z$ for dominant states at (B) $u_r=0.85$ (FM), (C)
				$u_r=0.90$ (PFM) and (D) $u_r=1.06$ (AFM).}\label{fig:2Dsim}
		\end{minipage}\hfill
		\begin{minipage}[t]{1\textwidth}
			\centering
			\includegraphics[width=1\textwidth]{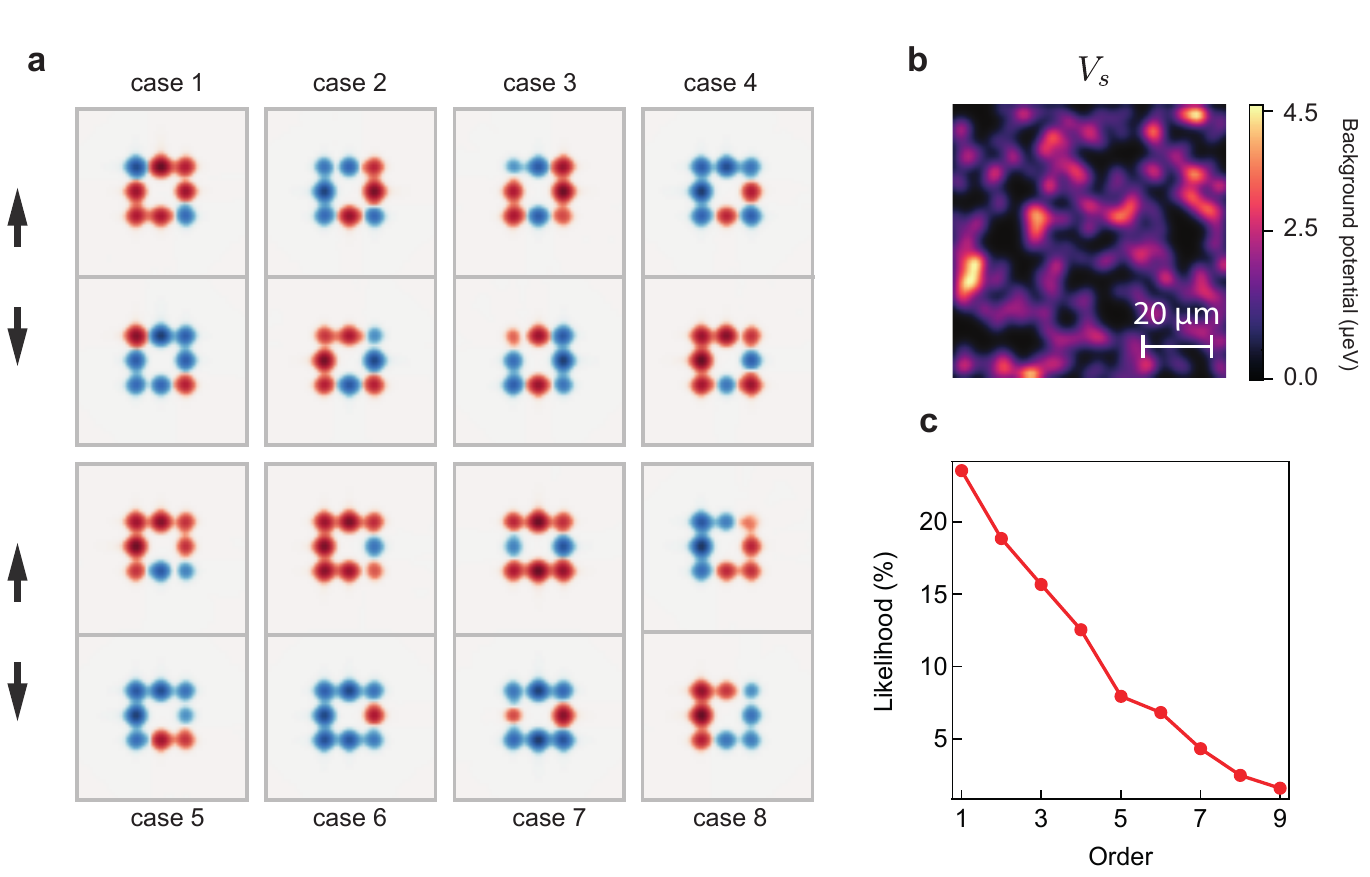}
			\caption{\captit{2D Simulations with random background potential}
				\capind{a} spin states at various cases of random background
				potential $V_s$. \capind{b} Typical case of a random background
				potential. \capind{c} Likelihood percentage of PCA orders for case 4.}\label{fig:2Dsimrandpot}
		\end{minipage}
	\end{figure*}

	For 2D simulations we use a complex Ginzburg-Landau-type
	equation~\cite{wouters_excitations_2007,
	keeling_spontaneous_2008} based on
	equation~\ref{eq:main}, which in addition incorporates the Laplacian, a
	repulsive potential due to the excitons in the pump spots, and energy
	relaxation~\cite{wouters_energy_2010} for polaritons in the trap:
	\begin{align}
		\label{eq:1}
		i\dot{\Psi}= & -\frac{i}{2} \left[ g(S)- \gamma\sigma_x \right]\Psi \nonumber \\
		             & +(1-i\Lambda)\left[
		\frac{1}{2}\left[(\alpha_1+\alpha_2)S+(\alpha_1-\alpha_2)S_z\sigma_z\right]\Psi
		- \frac{1}{2}\epsilon\sigma_x\Psi-\frac{\nabla^2}{2m^*}\Psi+(V_p+V_s)\Psi \right].
	\end{align}
	Here, $m^*$ is the effective mass of the polaritons and the harvest rate is
	given by $W=rP/\Gamma_R$, where $P$ is the spin-independent spatial profile
	of the excitation, $\Gamma_R$ is the decay rate of the exciton reservoir,
	and $r$ is the incoming rate of polaritons into the condensate.  The gain
	saturation is given by $\eta=r^2 P/\Gamma_R^2$, and it depends on the total
	occupation of the condensate (more generally treated in
	ref.~\citenum{ohadi_spontaneous_2015}). The repulsive potential due to the
	interaction of polaritons with the exciton reservoir is given by $V_p=\half
	g_r N + \half g_P P$, where $g_r$ and $g_P$ are the interaction constants of
	polaritons with the exciton reservoir and the pump spot respectively ($\hbar=1$), and
	$N=g(S)/r$ is the density of the exciton reservoir. Background potential due
	to inhomogeneity in the sample is given by $V_s$. Finally, $\Lambda\ll 1$ is
	a phenomenological constant that accounts for the energy relaxation, and it
	is proportional to the pump intensity.

    We initialize the coherent macroscopic wavefunction
    $\Psi$ with random noise and evolve it until the steady state is achieved. We
    repeat this process for 150 realizations and map out the correlation matrix of
    the condensate at each $u_r$.

	Fig.~\ref{fig:2Dsim} shows typical realizations at three distinct regimes of
	AFM, glass and FM.  For longer 8-condensate chains in the presence of a
	random background potential $V_s$, we observe the explicit appearance of
	glassy states as shown in Fig.~\ref{fig:2Dsimrandpot}. The random potentials
	in 2D simulations were acquired by adding randomly located Gaussian spots
	with $1/e$ diameters of \SI{12}{\um} and amplitudes of $0.8 g_P$.
Our 2D simulations show that a disorder potential of \SI{\sim5}{\ueV}
is enough to break spin chain symmetry.

 The stronger correlation in the experiment compared to simulations in
 Fig.~\ref{fig:3}(c,d) is due to the ramp-time of the pump laser. In the experiments, the
 rise-time is about 50 ns. In the model, the turn-on is instantaneous. The slow
 rise-time effectively drives the condensates adiabatically into the state with
 the lowest threshold state resulting in a highly deterministic outcome for each
 realization of the experiment. By contrast, if the pump turns on much faster
 than the dynamics of the condensate, the system can overshoot the lowest
 magnetization threshold, resulting in a more probabilistic outcome for each
 realization. In the simulations, the pump laser is turned on instantaneously at
 the start of the simulation and we find the steady state after \SI{2}{\nano\second}. These
 simulations are 2D and each point is the average of 150 realizations and
 calculating the whole graph takes about a day. In principle, one can
 incorporate the slow rise-time of the pump, but at the moment simulating
 adiabatic pumping unfortunately is beyond our current computational capacity.

The parameters used in the 2D simulations are
$\hbar\alpha_1=\SI{2}{\ueV\um^2}$; $\alpha_2=-0.5\alpha_1$; $\hbar
g_r=\SI{23}{\ueV\um^2}$; $g_P=g_R/4$; $\Lambda=0.1$;
$m^*=5.1\times10^{-5}m_e$; $r=\SI{0.05}{\ps^{-1}\um^2}$;
$\Gamma_R=\SI{10}{\ps^{-1}\um^2}$; $\Gamma=\SI{0.2}{\ps^{-1}\um^2}$;
$\hbar\epsilon=\SI{10}{\ueV}$; $\gamma=0.5 \epsilon$.

    \section{Qualitative analysis of steady spin states}
    \label{si:qualitative}
    \subsection{Mean-field model}

    The rate equations have a symmetry
    that can map the system onto a single condensate, using the ansatz:
    \begin{align}
            \begin{aligned}
            \Psi_{n+1} & =e^{i\varphi_{n+1,n}}\Psi_n, \qquad         & \text{\footnotesize (FM bond)}  \\
            \Psi_{n+1} & =e^{i\varphi_{n+1,n}}\sigma_x\Psi_n, \qquad & \text{\footnotesize (AFM bond)}
            \end{aligned}
    \end{align}
    where the phase-factors of each bond are to be determined. Making
    this ansatz, Eq.~\ref{eq:main} can be written as:
    \begin{align}
            \begin{split}
            i\dot{\Psi}_n = & -\frac{i}{2}(g(S_n)+i\omega_J)\Psi_n -\frac{i}{2}(\gamma -
            i\epsilon_{J})\sigma_x\Psi_n \\
                            & +\frac{1}{2}(\bar{\alpha}S_n+\alpha S_{nz}\sigma_z)\Psi_n.
            \end{split}
            \label{eq:dotpsi2}
    \end{align}
    This corresponds to a single condensate with a renormalized effective in-plane
    B-field $\epsilon_{J}$, arising from AFM bonds, and an energy-shift $\omega_J$
    arising from the FM bonds. The strength of these parameters depends on the
    relative phases between nearest neighbors in the system. For condensate $n$ one
    can write
    \begin{align}
            \begin{split}
            \epsilon_J & =\epsilon+J(\delta_{nk} e^{i\varphi_{nk}}+ \delta_{nl} e^{i\varphi_{nl}}) \\
            \omega_J   & = -J((1-\delta_{nk})e^{i\varphi_{nk}}+(1-\delta_{nl})e^{i\varphi_{nl}}),
            \end{split}
    \end{align}
    where $k,l$ are the nearest neighbors and $\delta_{nm} = 1,0$ for AFM or FM
    bonding respectively. Here $\varphi_{nm}$ is the phase shift moving from
    condensate $n$ to $m$. By mapping the system to a single mean-field condensate,
    we can take advantage of the single condensate analysis given in
    ref.~\citenum{ohadi_spontaneous_2015}, in particular the magnetization
    threshold $S_{c}=(\epsilon_J^2+\gamma^2)/\alpha\epsilon_J$, which for
    $\gamma<\epsilon$ is minimized when $\epsilon_J$ is minimized.

    \subsection{Criteria for state to dominate}
    \begin{enumerate}
    	\item {\bf The final state must be stable.} For a mean-field model, magnetization requires
    	      that
    	      $\epsilon_J>0$ (see ref.~\citenum{ohadi_spontaneous_2015}).
    	\item {\bf If multiple final states are stable, the state with the lowest spin
    		bifurcation threshold is favored.} Following turn-on of the pump, the
    		polariton number will increase, and the state that first reaches the
    		polariton number spin-bifurcation threshold is the most probable final
    		state. For a mean-field model, the critical spin-bifurcation threshold is
    		given by Eq. (4) of ref. \citenum{ohadi_spontaneous_2015}:
    		\begin{align}
    			S_{c}=\frac{\epsilon_J^2+\gamma^2}{\alpha\epsilon_J}. \label{eq:Sth}
    		\end{align}
    		Hence, the system favors states with the lowest $\epsilon_J$.
    	\end{enumerate}

    	\subsection{Application to 4 condensate system: stationary points}

    	\begin{figure*}
    		\centering
    		\begin{minipage}[t]{.48\textwidth}
    			\centering
    			\includegraphics[width=.9\textwidth]{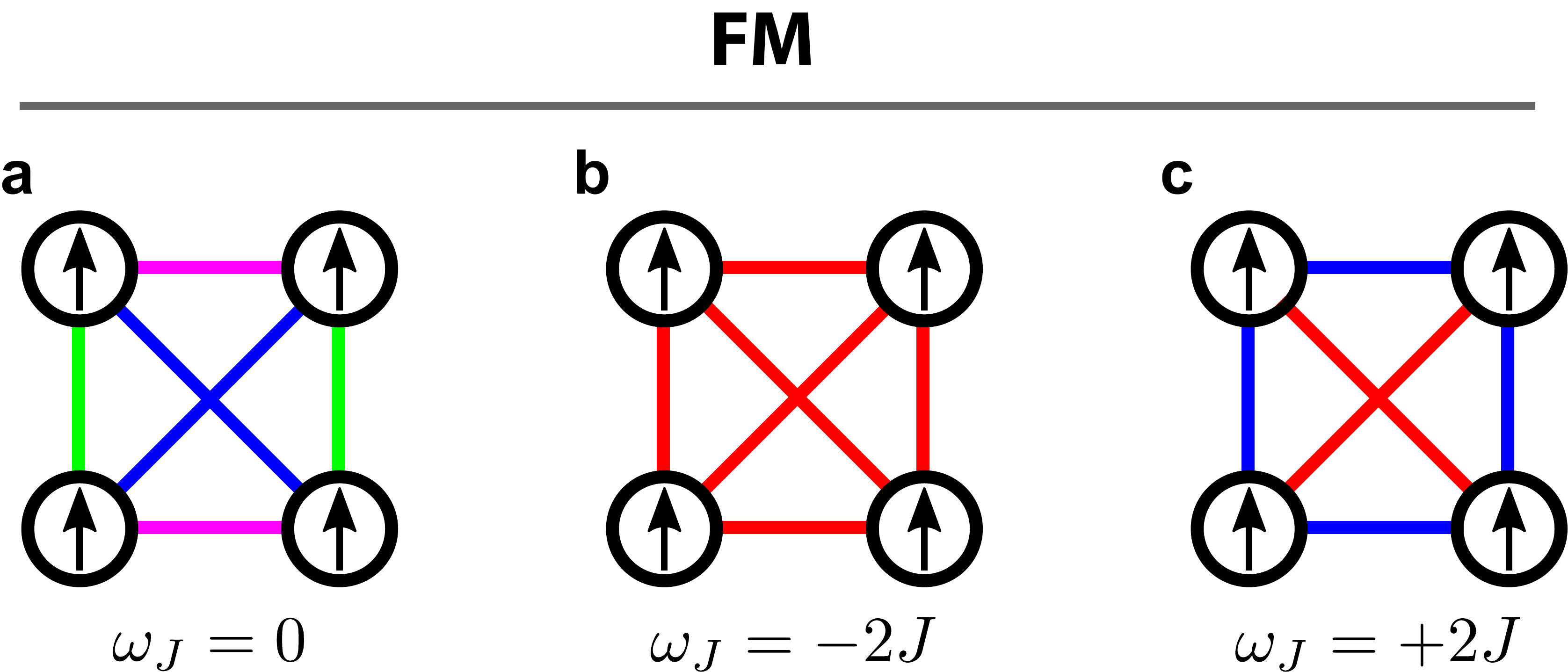} 
    			\caption{Schematic showing the three possible FM solutions. Red
    				lines show in-phase condensates ($\varphi_{nm} = 0$) whereas blue lines
    				are anti-phase ($\varphi_{nm} = \pi$). The pink and green
    				lines depict the arbitrary choice of phase between the nearest
    				neighbors as long as the condensates diagonally across are
    				anti-phase.}\label{fig.2x2_FMbonds}
    		\end{minipage}\hfill
    		\begin{minipage}[t]{.48\textwidth}
    			\centering
    			\centering
    			\includegraphics[width=0.9\textwidth]{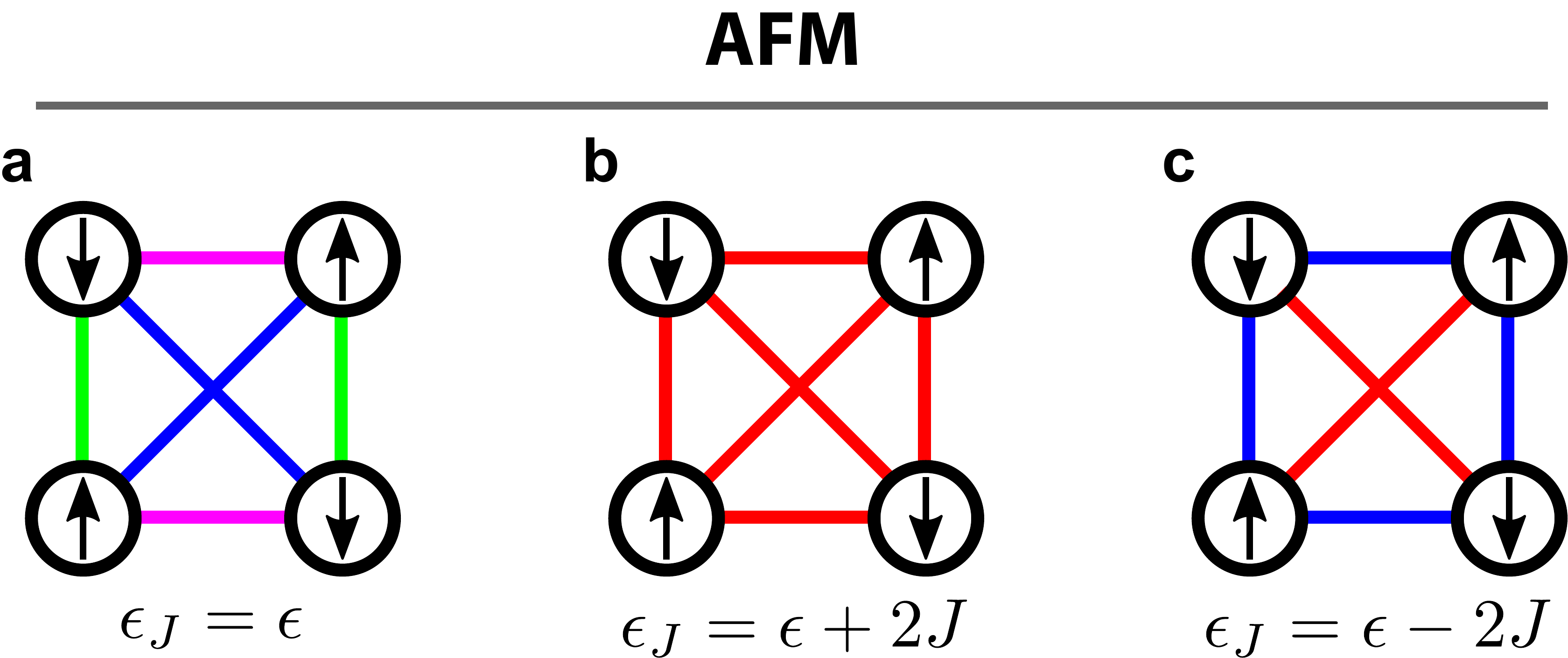} 
    			\caption{Schematic showing the three possible AFM solutions. Red
    				lines show in-phase condensates ($\varphi_{nm} = 0$) whereas blue lines
    				are anti-phase ($\varphi_{nm} = \pi$). The pink and green
    				lines depict the arbitrary choice of phase between the nearest
    				neighbors as long as the condensates diagonally across are
    				anti-phase.}\label{fig.2x2_AFMbonds}
    		\end{minipage}
    		\begin{minipage}{1\textwidth}
    			\centering
    			\includegraphics[width=0.55\textwidth]{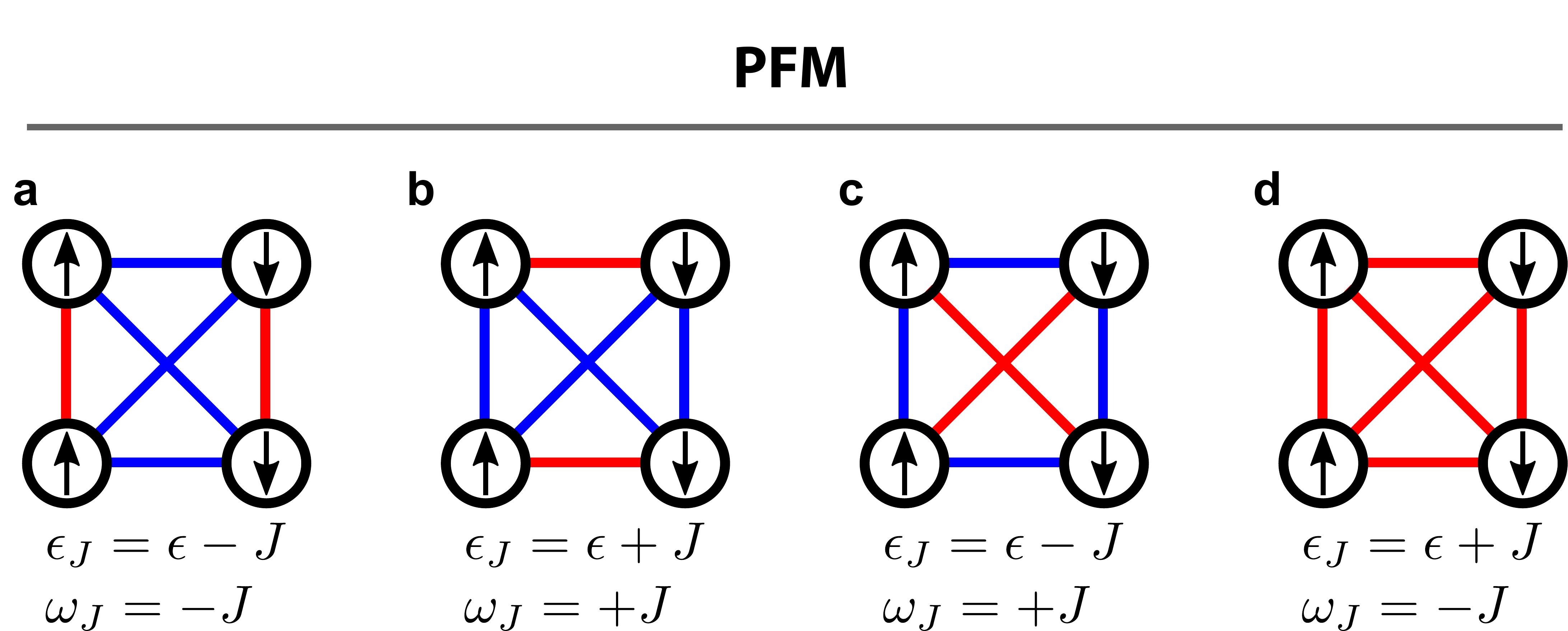} 
    			\caption{Schematic showing the four possible PFM solutions. Blue
    				lines are anti-phase condensates ($\varphi_{nm} = \pi$) and red are
    				in-phase ($\varphi_{nm} = 0$).}\label{fig.2x2_glassbonds}
    		\end{minipage}
    	\end{figure*}

    	To identify the stationary points of the system, we can use the Mean-field
    	model. We assume the stationary points are such that all condensates have
    	equal total polariton populations ($S_n$). This requires that both
    	$\epsilon_J$ and $\omega_J$ are real to avoid transferring population
    	between condensate sites, and are same for all sites. In principle, one can
    	have solutions where the magnitude of the spin is different for
    	different sites, but these solutions are highly nontrivial and are out of
    	the scope of the current work.

    	\subsubsection{FM stationary points}

    	For the FM state, since there are no AFM bonds, $\epsilon_J=\epsilon$. There are
    	three possible ways to make $\omega_J=-J(e^{i\varphi_{nk}}+e^{i\varphi_{nl}})$
    	real and constant across all sites to have all 4 condensates phase locked:
    	\begin{enumerate}[noitemsep]
    		\item The diagonal states are in anti-phase, $\omega_J=0$ (Fig.~\ref{fig.2x2_FMbonds}a).
    		\item Nearest neighbors are in-phase,
    		      $\omega_J=-2J$ (Fig.~\ref{fig.2x2_FMbonds}b)
    		\item Nearest neighbors are anti-phase,
    		      $\omega_J=+2J$ (Fig.~\ref{fig.2x2_FMbonds}c)
    	\end{enumerate}
    	Since all states have the same $\epsilon_J$, they share the same spin-bifurcation threshold.

    	\subsubsection{AFM stationary points}

    	For the AFM state, since there are no FM bonds, $\omega_J=0$. Similar to the FM
    	state, there are three possible ways to make
    	$\epsilon_J=-J(e^{i\varphi_{nk}}+e^{i\varphi_{nl}})$ real, and
    	constant across all sites:
    	\begin{enumerate}[noitemsep]
    		\item[1.] Sites diagonally across are anti-phase, $\epsilon_J=\epsilon$
    		      is unchanged (Fig.~\ref{fig.2x2_AFMbonds}a).
    		\item[2.] All nearest neighbors are in-phase, $\epsilon_J =\epsilon+2J$
    		      (Fig.~\ref{fig.2x2_AFMbonds}b).
    		\item[3.] All nearest neighbors are anti-phase,
    		      $\epsilon_J=\epsilon-2J$ (Fig.~\ref{fig.2x2_AFMbonds}c).
    	\end{enumerate}

    	By inspection of Eq.~\ref{eq:Sth}, AFM state (3) with anti-phase nearest
    	neighbors has the smallest $\epsilon_J$, and consequently the lowest spin
    	bifurcation threshold of the possible AFM states.

    	\subsubsection{PFM stationary points}

    	For the PFM state, the system is characterized by each condensate having
    	one FM bond and AFM bond. There are four possible combinations which are
    	given in Fig.~\ref{fig.2x2_glassbonds} which allow $\omega_J$ and
    	$\epsilon_J$ to stay real. From the analysis above, one can easily see that
    	the solutions are combinations of $\epsilon_J = \epsilon \pm J$ and
    	$\omega_J = \pm J$ depending on whether the bonds are anti-phase or
    	in-phase. The minimum bifurcation threshold then selects the state with an
    	anti-phase AFM bond, such that $\epsilon_J=\epsilon-J$.

    	\subsection{Mean-field phase diagram}

    	\begin{figure*}
    		\begin{minipage}[t]{.48\textwidth}
    			\centering
    			\includegraphics[width=0.9\linewidth]{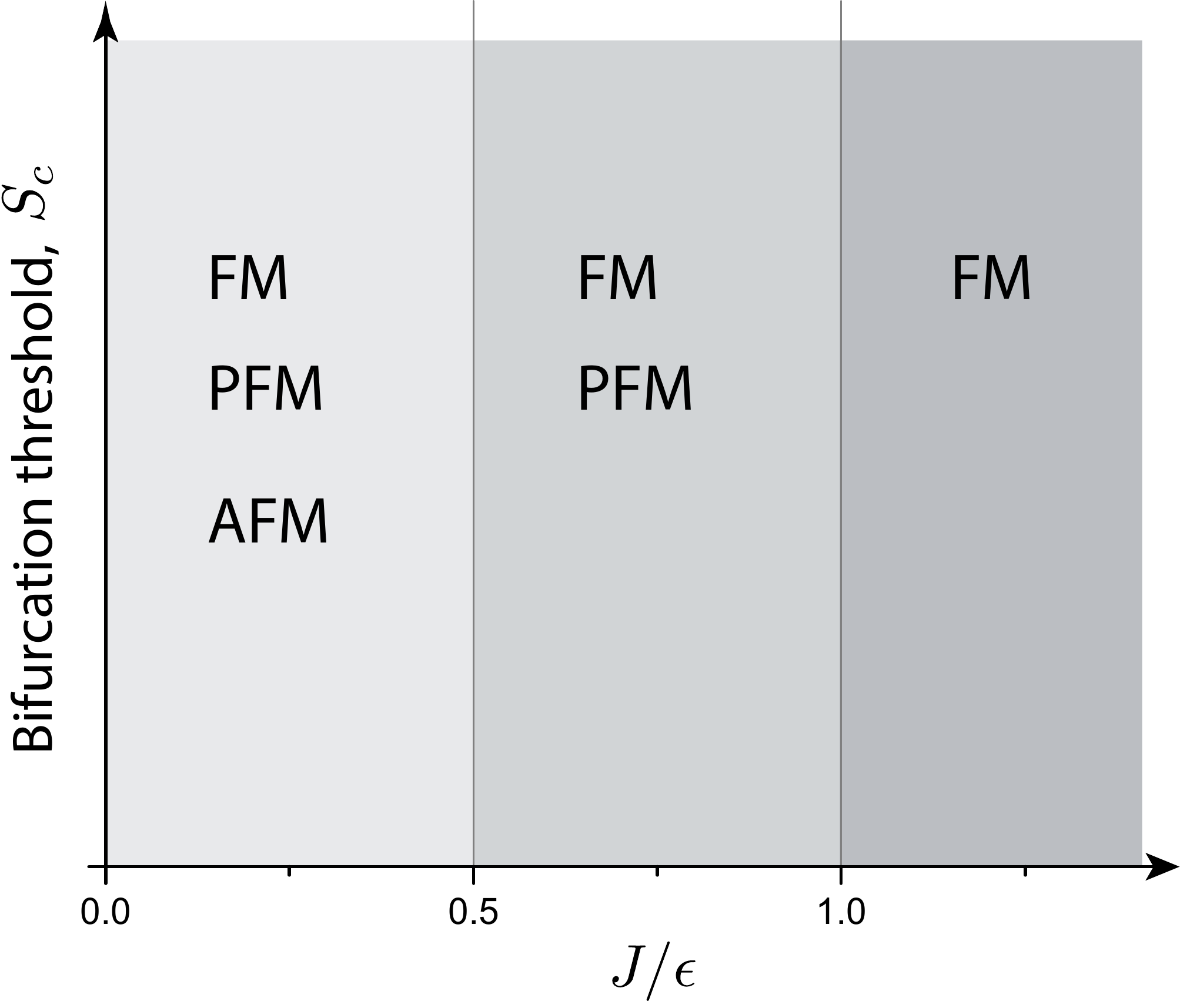}
    			\caption{\captit{Phase-diagram of 4 condensate system using a mean-field model.} For
    				small $J/\epsilon< 0.5$, all states are potentially stable. However the AFM
    				state wins because it has the lowest spin bifurcation threshold. The AFM
    				state becomes unstable for $J/\epsilon>0.5$, and the PFM state is observed
    				because it is now the state with the lowest threshold. For $J/\epsilon>1$,
    				only the FM state is stable.}\label{fig:4condphasediag}
    		\end{minipage}\hfill
    		\begin{minipage}[t]{.48\textwidth}
    			\centering
    			\includegraphics[width=.8\linewidth]{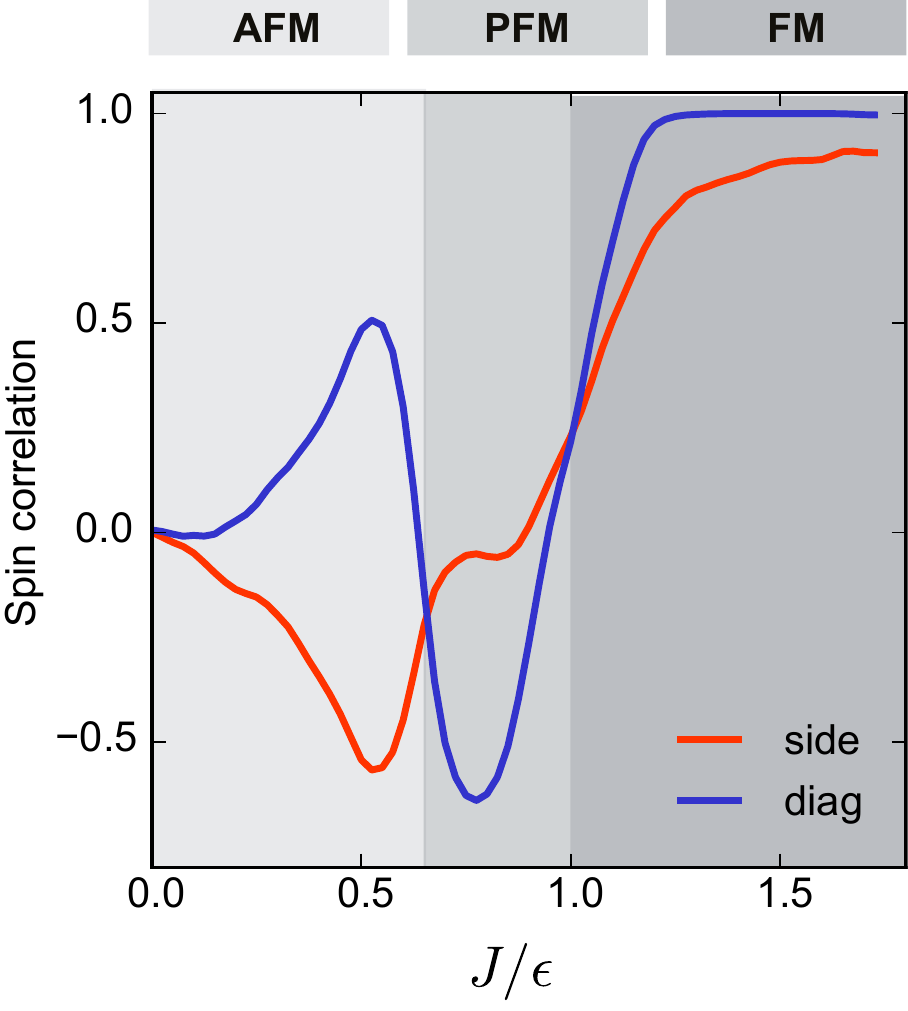}
    			\caption{\captit{0D simulations} Monte-Carlo simulations of the 0D model showing spin correlation of
    				the side and diagonal condensates in a 4-condensate chain as a function
    				of $J/\epsilon$.}\label{fig:0D}
    		\end{minipage}
    	\end{figure*}

    	The spin-bifurcation thresholds can be summarized as follows: $\epsilon_J >
    	0$ for the solution to exist but the smallest $\epsilon_J$ has the lowest
    	threshold (see  Eq.~\ref{eq:Sth}). Considering thus only solutions where
    	$\epsilon_J$ is minimal, the coupling parameter $J$ defines a boundary
    	between AFM, PFM, and FM solutions which have two, one, and zero
    	(anti-phase) AFM bonds respectively:
    	\begin{align}
    		\begin{split}
    		  & \epsilon_{FM}=\epsilon,      \\
    		  & \epsilon_{PFM}=\epsilon-J, \\
    		  & \epsilon_{AFM}=\epsilon-2J.
    		\end{split}
    	\end{align}
    	Hence, for $J<\epsilon/2$, all spin states are stable, and the AFM state is
    	observed since it has the lowest spin-bifurcation threshold. For
    	$\epsilon/2<J<\epsilon$, the AFM state is no longer stable, and the PFM
    	state is observed since it has the lowest threshold. For $J>\epsilon$, only
    	the FM states are stable, as shown in Fig.~\ref{fig:4condphasediag}.

\end{document}